\documentclass[prx,twocolumn]{revtex4-2}
\usepackage{amsmath, amssymb, mathtools, hyperref}
\usepackage{amsthm}
\usepackage{color}
\usepackage{float}

\usepackage[explicit]{titlesec}
\titlespacing{\section}{0pt}{2ex}{1ex}
\titlespacing{\subsection}{0pt}{1ex}{0ex}
\titlespacing{\subsubsection}{0pt}{0.5ex}{0ex}

\def\eu{\ensuremath{\mathrm{e}}}
\def\iu{\ensuremath{\mathrm{i}}}
\def\du{\ensuremath{\mathrm{d}}}

\newtheorem{approximation}{Approximation}
\newtheorem{assumption}{Condition}
\newtheorem{effectivemodel}{Closure Model}

\global\long\def\argmin{\operatornamewithlimits{argmin}}

\begin{document}

\title{Data-driven Selection of Coarse-Grained Models of Coupled Oscillators}
\author{Jordan Snyder$^{1,2}$, Anatoly Zlotnik$^3$, Andrey Y. Lokhov$^3$}
\affiliation{$^1$Department of Mathematics, University of California, Davis, CA 95616}
\affiliation{$^2$Department of Applied Mathematics, University of Washington, Seattle, WA 98195}
\affiliation{$^3$Theoretical Division, Los Alamos National Laboratory, Los Alamos, NM 87545}

\begin{abstract}
Systematic discovery of reduced-order closure models for multi-scale processes remains an important open problem in complex dynamical systems. Even when an effective lower-dimensional representation exists, reduced models are difficult to obtain using solely analytical methods. Rigorous methodologies for finding such coarse-grained representations of multi-scale phenomena would enable accelerated computational simulations and provide fundamental insights into the complex dynamics of interest. We focus on a heterogeneous population of oscillators of Kuramoto type as a canonical model of complex dynamics, and develop a data-driven approach for inferring its coarse-grained description. Our method is based on a numerical optimization of the coefficients in a general equation of motion informed by analytical derivations in the thermodynamic limit. We show that certain assumptions are required to obtain an autonomous coarse-grained equation of motion. However, optimizing coefficient values enables coarse-grained models with conceptually disparate functional forms, yet comparable quality of representation, to provide accurate reduced-order descriptions of the underlying system.
\end{abstract}

\maketitle

\section{Introduction}

Numerical simulations of complex multi-scale phenomena are fundamental to modern science, for which commonly sought goals involve the development of tractable yet accurate reduced-order models. There are many approaches to this sort of problem throughout many diverse domains. For instance, in turbulence modeling, this is known as the closure problem for the Reynolds-Averaged Navier-Stokes equation (RANS) \cite{Reynolds1895, Chou1945, Pope1975, Duraisamy2019}. In molecular dynamics, it is known simply as coarse-graining \cite{Foley2015,Boninsegna2018,Wang2019}. Mean-field approaches to analyzing stochastic dynamics on networks also fall into this description \cite{Gleeson2012,Porter2016}.

Investigations on model reduction of nonlinear dynamics often consider the Kuramoto model of coupled oscillators \cite{kuramoto1975self}, which has long been studied as an example of collective behavior \cite{aschoff1981circadian,Strogatz1989,Frank2000,Arenas2006,Li2008}. The key features of the Kuramoto model are its composition of many oscillatory units with distinct natural frequencies, and pairwise coupling that tends to drive phases together. Its popularity as an object of study comes from its tractability in certain special limits \cite{strogatz2000kuramoto,ott2008low} and its nonetheless rich phenomenology \cite{Panaggio2015,Bick2018,Zhang2019b}. In addition, the Kuramoto model and its variants describe a number of synchronization phenomena in domains of diverse nature such as coupled Josephson junctions \cite{wiesenfeld1998frequency}, neuroscience \cite{varela2001brainweb}, chemical oscillators \cite{kiss2002emerging,zlotnik2016phase}, and the power grid \cite{dorfler2013synchronization}.  

Substantial work on the Kuramoto model focuses on understanding synchronization under various conditions on the natural frequencies and on the structure of the network that couples oscillators to one another. One main mode of understanding consists of finding simplified mathematical descriptions of the dynamics, and this is where the closure problem arises. Perhaps the best-known example of this type is the seminal work by Ott and Antonsen \cite{ott2008low}, who showed that in the $N\to\infty$ limit and under certain conditions on the distribution of natural frequencies and initial phases, the center of mass of a population of Kuramoto oscillators obeys an autonomous ODE. Other studies have focused more on complex coupling topologies, proposing techniques using spectral information to merge nodes together \cite{Gfeller2008} or otherwise systematically discard irrelevant degrees of freedom \cite{Izumida2013}. Still others take a more strictly data-driven approach and seek e.g. closed equations of motion for low-order moments of the distribution of phases \cite{Moon2006,Rajendran2011} or to identify good coarse-grained variables via manifold learning techniques \cite{Thiem2020}. Finally there are approaches that employ a ``collective coordinate'' ansatz governing the phase of each oscillator within a phase-locked cluster, and thereby arrive at a closed equation of motion \cite{Gottwald2015,Hancock2018,Smith2019,Smith2019a,Yue2020}.

Here we use a collective coordinate ansatz to derive coarse-grained equations of motion at the level of phase-locked clusters that are consistent with arbitrary distributions of natural frequency within each cluster. We focus especially Gaussian and Cauchy distributions, though our approach is generic. The resulting equations are precisely determined in the $N\to\infty$ limit in terms of parameters of the distributions in question and the matrix of coupling strengths. Similar to other previously obtained analytical results, the emergence of an explicit closure model crucially depends on a number of simplifying assumptions, and no closed-form reduced-order equations of motion are known for finite systems that do not satisfy these assumptions. We aim to move beyond the $N\to\infty$ limit by treating our derived coarse-grained equations as inductive biases, and allowing data from finite-$N$ simulations to determine optimal parameter values. The result is a systematic data-driven procedure for finding physically meaningful coarse-grained models of finite systems of coupled oscillators, that provide a more accurate reduced-order description of the system.

This manuscript is organized as follows. In Sec. \ref{sec:background} we review in detail prior work and precisely state the problem we address. In Sec. \ref{subsec:assumptions} we state the set of conditions and approximations that let us derive coarse-grained equations of motion for coupled oscillator systems, and in Sec. \ref{sec:inference} we introduce our data-driven approach for finding optimal coefficient values to use in these coarse-grained equations. In Sec. \ref{sec:results} we apply our methods to a concrete example system and evaluate the performance of both our theoretically-derived coarse-grained models, and the same models optimized to fit training data.

\section{Background} \label{sec:background}

\noindent The Kuramoto model is the ordinary differential equation (ODE) system
\begin{equation}
\dot{\theta}_i = \omega_i + \sum_{j=1}^N K_{ij}\sin(\theta_j - \theta_i),\qquad i=1\dots N,
\label{eq:kuramoto}
\end{equation}
where $\theta_i\in S^1$ is the \emph{phase} of the $i^\textrm{th}$ oscillator, $\omega_i\in \mathbb{R}$ is its \emph{natural frequency}, $N$ is the total number of oscillators, and $K\in \mathbb{R}^{N\times N}$ is the \emph{coupling matrix} that defines which oscillators influence each other \cite{kuramoto1975self}.

We can equivalently formulate the above model in terms of \emph{complex phases} $y_i = \exp(\iu \theta_i)$. A straightforward calculation yields the equivalent representation
\begin{equation}
    \dot{y}_i = \iu \omega_i y_i + \frac{1}{2} \sum_{j=1}^N K_{ij}(y_j -  y_j^* y_i^2),
    \label{eq:kuramoto_complex}
\end{equation}
where we have used the fact that $y_i^*y_i = 1$. We subsequently explain why this representation of the Kuramoto dynamics is more convenient for our purposes.

Perhaps the best-studied case of the Kuramoto model is that of \emph{mean-field} coupling, where $K_{ij} = K/N$ for all $i,j$. In this case {it is well known that in the limit $N\to\infty$, the system \eqref{eq:kuramoto} exhibits a phase transition with respect to $K$}: if $\omega_i$ are sampled from a symmetric, unimodal distribution, then there exists $K_c$ such that for $K<K_c$, the oscillators behave mostly independently, while for $K>K_c$ a subset of oscillators spontaneously locks to a single frequency \cite{Kuramoto1984}.

Synchronization in the Kuramoto model is typically quantified by the order parameter,
\begin{equation}
z \coloneqq \frac{1}{N}\sum_{j=1}^N \eu^{\iu \theta_j} = R\eu^{\iu \Phi},
\label{eq:global_order_parameter}
\end{equation}
where $R\in[0,1]$ is the \emph{synchrony} and $\Phi\in[0,2\pi)$ is the \emph{average phase}. If all phases are equal then $R=1$, and if the phases are spread uniformly over the unit circle, then $R\approx 0$. Thus $R$ is a natural measure of synchronization.

The dynamics of $z$ depend on the dynamics of all $\theta_i$, but it is natural to suppose that in some limit there exists a \emph{closed} equation for the dynamics of $z$. Indeed there is, as demonstrated by Ott and Antonsen \cite{ott2008low}. Assuming that $\omega_i$ are Cauchy-distributed, i.e., $\omega_i \sim g(\omega)$ where $g$ is the Cauchy probability density function with mode $\Omega$ and width $\delta$, and that the coupling is mean-field, then $z$ evolves according to a Stuart-Landau equation,
\begin{equation}
    \dot{z} = \left(\iu \Omega - \delta + {K\over 2}\right) z - {K \over 2} z |z|^2.
    \label{eq:stuart-landau-one-var}
\end{equation}
The form \eqref{eq:stuart-landau-one-var} shows clearly that $z=0$ is always a solution, but undergoes a pitchfork bifurcation at $K_c = 2\delta$, when a new solution with $R = \sqrt{1-2\delta/K}$ (and $\du \Phi/\du t = \Omega$) appears, representing partial synchrony that becomes global synchrony (i.e. $R\to 1$) as $K\to \infty$.

Interestingly, the same analysis carries over to the case where oscillators are not coupled all-to-all, but are divided into subsets such that the strength of coupling between any two oscillators depends on the subsets to which they belong. Let $\Pi = (P_1, \dots, P_C)$ be a partition of the index set $\{1, \dots, N\}$ and let $K$ be a $C\times C$ matrix of coupling strengths. Following \cite{ott2008low}, such a modular system can be written as 
\begin{equation}
\dot{\theta}_i = \omega_i + \sum_{\sigma' = 1}^{C}  \frac{K_{\sigma \sigma'}}{|P_{\sigma'}|} \cdot \sum_{j \in P_{\sigma'}} \sin(\theta_j - \theta_i)
\label{eq:kuramoto_modular}.
\end{equation}

The same mathematical machinery as before can be applied to show that if for every $\sigma$, $\{\omega_i | i\in P_\sigma\}$ are distributed according to a Cauchy distribution with mode $\Omega_\sigma$ and width $\delta_\sigma$, then the \emph{cluster order parameters} $\{z_\sigma\}$, defined by
\begin{equation}
    z_\sigma = \left\langle \eu^{\iu \theta_j} \right\rangle_{j\in P_\sigma} \coloneqq \frac{1}{|P_\sigma|} \sum_{j\in P_\sigma} \eu^{\iu \theta_j}, 
    \label{eq:local_order_params}
\end{equation}
obey a coupled Stuart-Landau equation of the form
\begin{equation}
\dot{z}_\sigma = \iu (\Omega_\sigma + \iu \delta_\sigma)z_\sigma +\frac{1}{2}\sum_{\sigma' = 1}^C K_{\sigma \sigma'} \left(z_{\sigma'} - z_{\sigma'}^* z_\sigma^2\right).
\label{eq:skardal_restrepo_CG_model}
\end{equation}
{The consequences of this equation for the existence of mesoscale synchronization were discussed in detail in a prior study \cite{Skardal2012}. For now, we note that the above representation is mathematically equivalent to the original Kuramoto model written in terms of complex phases (i.e. Eq. \eqref{eq:kuramoto_complex}), albeit with the linear term containing a nonzero real part.} In this sense, this natural emergence of Eq. \eqref{eq:skardal_restrepo_CG_model} implies that, in the Kuramoto model, groups of oscillators can behave collectively as a single oscillator.  In other words, there is a renormalization procedure for coupled oscillator systems that remains within the same model class (in particular, the one defined by Eq. \eqref{eq:skardal_restrepo_CG_model}).

The above analyses are valid in the limit $N\to\infty$, and require certain regularity assumptions on the initial distribution of phases. As such, the behavior of a finite-$N$ Kuramoto system will in general exhibit fluctuations relative to \eqref{eq:stuart-landau-one-var} or \eqref{eq:skardal_restrepo_CG_model}. On the other hand, the dynamics of any finite-$N$ system in the absence of noise are in fact deterministic, and hence its evolution can be computed exactly from its current ($N$-dimensional) state. This raises the natural question: Given a Kuramoto system of the form \eqref{eq:kuramoto}, what is the coarsest partition $\Pi = (P_1, \dots, P_C)$ such that the local order parameters (\ref{eq:local_order_params})
evolve according to an autonomous ODE? And, what is that ODE?

Alternatively, we suppose that $\Pi$ is a partition and $z^\Pi = (z_1, \dots, z_C)\in \mathbb{C}^C$ is the vector of local order parameters as defined above. Does the value of $z^\Pi$ exactly determine the value of $\dot{z}^\Pi$? And if so, what is the functional relationship $\dot{z}^\Pi = f(z^\Pi)$? Our proposal to address this question is to learn the function $f$ from samples $\{(z^\Pi(t), \dot{z}^\Pi(t)) | t \in [0, T] \}$ obtained by coarse-graining experimental or simulated dynamics data of the full system. This learning task becomes a well-posed optimization problem when a finite set of basis functions that can be used to construct $f$ is chosen. Finding $f$ such that $\Vert \dot{z}^\Pi - f(z^\Pi) \Vert$ is small means that the dynamics of $z^\Pi$ (i.e. $\dot{z}^\Pi$) can be accurately computed from only the value of $z^\Pi$, and we say that the system admits a good coarse-grained model described by $f(z^\Pi)$.

\section{Methods} \label{sec:methods}

We now propose criteria to select the coarse-graining partition $\Pi$, and present analytical calculations that suggest which finite-dimensional function spaces are likely to contain a satisfactory coarse-grained model in the sense of approximating cluster-average phase trajectories. We do this by imposing certain assumptions on the distribution of individual oscillators' phases, and arrive at expressions for the coarse-grained model in terms of static parameters of the fine-grained model. These expressions hence suggest a natural set of \emph{basis functions} that can be used to inform our data-driven procedure for inferring coarse-grained dynamics directly from observed time-series data.

\subsection{Conditions and Analytical Implications} \label{subsec:assumptions}

To fix notation, we consider the Kuramoto model in complex form (\ref{eq:kuramoto_complex}), and refer interchangeably to either real-valued phases $\theta_i$ or their complex versions $y_i \coloneqq \exp(\iu \theta_i)$.
It is possible to write exact evolution equations for the coarse variables $z_\sigma$ (\ref{eq:local_order_params}) if we allow explicit time-dependence through the residuals $x_i(t) = z_\sigma(t) - y_i(t)$. Explicitly, we have
\begin{align}
   \!\! \dot{z}_\sigma &= A_\sigma + {B_\sigma z_\sigma} + D_\sigma z_\sigma^2 + \nonumber \\
    & + \sum_{\sigma' = 1}^C \left[ {E_{\sigma \sigma'}(z_{\sigma'} - z_{\sigma'}^* z_\sigma^2)} + F_{\sigma \sigma'}z_{\sigma'}^* + G_{\sigma \sigma'}z_{\sigma}z_{\sigma'}^* \right]
    \label{eq:reduced_eqn_terms_gathered}
\end{align}
where each coefficient ($A$)--($G$) is a function of the residuals $x_i(t)$. The terms with coefficients $B_\sigma$ and $E_{\sigma \sigma'}$ are those that are present in the original Kuramoto equations written in complex form.

The functional form of the coefficients is given by
\begin{align}
    A_\sigma &= \left\langle \iu \omega_i x_i + \sum_{\sigma'=1}^C \sum_{j\in P_{\sigma'}} K_{ij} (x_j - x_j^* x_i^2)\right\rangle_{i\in P_\sigma} \label{eq:A}\\
    B_\sigma &= \left\langle \iu \omega_i + \sum_{\sigma'=1}^C \sum_{j\in P_{\sigma'}} K_{ij} (-2x_j^* x_i)\right\rangle_{i\in P_\sigma}\label{eq:B}\\
    D_\sigma &= \left\langle \sum_{\sigma'=1}^C \sum_{j\in P_{\sigma'}} K_{ij} (-x_j^*)\right\rangle_{i\in P_\sigma}\label{eq:D}\\
    E_{\sigma \sigma'} &= \left\langle \sum_{j\in P_{\sigma'}} K_{ij}\right\rangle_{i\in P_\sigma} \label{eq:E}\\
    F_{\sigma \sigma'} &= \left\langle \sum_{j\in P_{\sigma'}} K_{ij} (-x_i^2)\right\rangle_{i\in P_\sigma}\label{eq:F}\\
    G_{\sigma \sigma'} &= \left\langle \sum_{j\in P_{\sigma'}} K_{ij} (-2x_i)\right\rangle_{i\in P_\sigma} \label{eq:G}
\end{align}

If one could obtain an accurate expression for $x_i$ in terms of microscopic parameters $\omega_i$, $K_{ij}$, and the coarse variables $z_\sigma$, then the explicit time-dependence could be removed, leaving an autonomous ODE for the coarse variables $\{z_\sigma\}$. In order to arrive to simplified expressions, we build on the work of Gottwald concerning collective coordinates \cite{Gottwald2015}. That study proposes that the residual phase of each oscillator within a phase-locked cluster is directly proportional to its natural frequency relative to the cluster average frequency. The associated constant of proportionality serves as a ``collective coordinate'', which enables an expression involving all phase variables in terms of the cluster average phase and the collective coordinate. Accordingly, we describe how to simplify the above equations based on conditions for a collective coordinate ansatz to be appropriate.

\begin{assumption} \label{ass:modularity-of-coupling-matrix}
{\bf Modularity of the coupling matrix}. The coupling matrix $K$ is such that $K_{ij} = K_{\sigma \sigma'}$ whenever $i\in P_{\sigma}$ and $j\in P_{\sigma'}$.
\end{assumption}

Through Condition \ref{ass:modularity-of-coupling-matrix}, we require that the coupling strength between any two oscillators is a function only of their module memberships. Importantly, Condition \ref{ass:modularity-of-coupling-matrix} is preserved under refinement; if $\Pi_1$ refines $\Pi_2$ and $\Pi_2$ satisfies Condition \ref{ass:modularity-of-coupling-matrix}, then $\Pi_1$ also satisfies Condition \ref{ass:modularity-of-coupling-matrix}. It is also true that there is a unique coarsest partition $\Pi^\text{struct}$ that satisfies Condition \ref{ass:modularity-of-coupling-matrix}, and any other partition satisfying Condition \ref{ass:modularity-of-coupling-matrix} refines $\Pi^\text{struct}$; to see this, note that $\Pi^\text{struct}$ can be constructed as the set of equivalence classes under the relation $i \equiv j \iff  K_{i\cdot} = K_{j\cdot} \wedge K_{\cdot i} = K_{\cdot j}$, i.e. two nodes are equivalent if and only if their in- and out-neighborhoods are identical. We refer to $\Pi^\text{struct}$ as the \emph{structural} partition. Note that this definition is more stringent than other partitions that have been used elsewhere such as (external) equitable partitions \cite{Schaub2016} or orbit partitions induced by the automorphism group of the coupling matrix \cite{Pecora2014, Sorrentino2015}. Condition \ref{ass:modularity-of-coupling-matrix} greatly simplifies several of the coefficients; we have
\begin{align}
    A_\sigma &= \left\langle \iu \omega_i x_i\right\rangle_{i\in P_\sigma} \label{eq:A_simp}\\
    B_\sigma &= \left\langle \iu \omega_i \right\rangle_{i\in P_\sigma}\label{eq:B_simp}\\
    D_\sigma &= 0\label{eq:D_simp}\\
    E_{\sigma \sigma'} &= K_{\sigma \sigma'}|P_{\sigma'}| \label{eq:E_simp}\\
    F_{\sigma \sigma'} &= - K_{\sigma \sigma'} |P_{\sigma'}| \left\langle x_i^2\right\rangle_{i\in P_\sigma}\label{eq:F_simp}\\
    G_{\sigma \sigma'} &= 0 \label{eq:G_simp}
\end{align}

Notice that the dependence on the fine-grained state of the system now appears only in (\ref{eq:A_simp}) and (\ref{eq:F_simp}), so with an appropriate ansatz for the $\{x_i\}$ in terms of $\{z_\sigma\}$ and parameters, a closed-form system can be obtained.

\begin{assumption} \label{ass:phase-cohesive-clusters}
{\bf Phase-cohesiveness of clusters}. Each cluster is phase-cohesive, meaning that $|\theta_i(t) - \theta_j(t)|$ remains bounded for all $t$, whenever oscillators $i$ and $j$ belong to the same cluster.
\end{assumption}

As before, Condition \ref{ass:phase-cohesive-clusters} is a property preserved under refinement, and the set of partitions that satisfy it can be defined by a coarsest partition $\Pi^\text{dyn}$, the \emph{dynamical} partition. Any partition satisfying Condition \ref{ass:phase-cohesive-clusters} is a refinement of $\Pi^\text{dyn}$, and one can obtain $\Pi^\text{dyn}$ as the set of equivalence classes under the equivalence relation 
\begin{equation}
   \!\! i \equiv j \! \iff \!\! \lim_{T\to\infty}{\theta_i(T) - \theta_i(0) \over T} \!=\! \lim_{T\to\infty}{\theta_j(T) - \theta_j(0) \over T}
   \label{eq:dynamical_equivalence_relation}
\end{equation}
This enables the following approximation, which underpins the present derivation.

\begin{approximation} \label{app:linear-collective-coordinate}
{\bf Linear collective coordinate ansatz}. We take a simple collective coordinate ansatz, namely that the phase residual of each oscillator around its cluster mean is directly proportional to its frequency residual about its cluster mean. Formally, we assume
\begin{equation}
    \theta_i = \arg(z_\sigma) + \alpha_\sigma\tilde{\omega}_i
    \label{eq:collective_coord_ansatz}
\end{equation}
where $\tilde{\omega}_i = \omega_i - \langle \omega_i \rangle_{i\in P_\sigma}$ is the residual of oscillator $i$'s natural frequency relative to the mean in its cluster and $\alpha_\sigma$ is a cluster-dependent constant of proportionality, to be determined.
\end{approximation}

Note that this is only one choice of many; another natural choice is that the phase residual depends not linearly on the frequency residual but on its arcisne. We also note that the Conditions \ref{ass:modularity-of-coupling-matrix} and \ref{ass:phase-cohesive-clusters} are necessary for such an ansatz to be appropriate, since under these Conditions the only remaining distinction between oscillators within a cluster is in their natural frequencies. Thus we may reasonably expect that the dynamics of different oscillators in a cluster differ in a regular way.

The proportionality constant $\alpha_\sigma$ must be such that $\langle y_i \rangle_{i\in P_\sigma} = z_\sigma$, and it is this constraint that lets us finally obtain an autonomous ODE for $\{z_\sigma\}$.
This leads to a consistency condition that reads
\begin{align}
    \langle y_i \rangle_{i\in P_\sigma} = \frac{z_\sigma}{|z_\sigma|} \chi_\sigma(\alpha_\sigma),
    \label{eq:consistency-relation-for-alpha}
\end{align}
where $\chi_\sigma$ is the characteristic function of the distribution of natural frequency residuals $\tilde{\omega}_i = \omega_i - \langle \omega_i \rangle_{i\in P_\sigma}$ within cluster $\sigma$. This condition gives the value of $\alpha_\sigma$ implicitly in terms of $z_\sigma$, and thereby lets us describe the entire distribution of phases within a given cluster - and hence the expectations in Eqs. (\ref{eq:A_simp}) and (\ref{eq:F_simp}) - in terms of the single variable $z_\sigma$. The general formulae are
\begin{equation}
\begin{aligned}
    \langle \iu \omega_i x_i \rangle_{i\in P_\sigma} &= \frac{z_\sigma}{|z_\sigma|} \chi_\sigma'(\chi_\sigma^{-1}(|z_\sigma|))\\
    \langle x_i^2 \rangle_{i\in P_\sigma} &= z_\sigma^2 \left({1 \over |z_\sigma|^2} \chi_\sigma(2\chi_\sigma^{-1}(|z_\sigma|) - 1 \right).
    \label{eq:general_cc_closure_terms}
\end{aligned}    
\end{equation}

If for, example, the natural frequency residuals are distributed according to a Gaussian distribution with variance $v_\sigma$, then $\chi_\sigma(\alpha_\sigma) = \exp(-v_\sigma\alpha_\sigma^2 / 2)$, and
we have
\begin{align}
    \langle\iu \omega_i x_i \rangle &= -z_\sigma ( \sqrt{-2v_\sigma\ln(|z_\sigma|)}) \label{eq:A-closure-GCC} \\
    \langle x_i^2 \rangle &= z_\sigma^2 (|z_\sigma|^2 - 1). \label{eq:F-closure-GCC}
\end{align}
Substituting the above expressions into Eq. \eqref{eq:reduced_eqn_terms_gathered} yields a closed system of equations for the dynamics of the coarse-grained variables, whose parameters are directly computable from the parameters of the fine-grained system.

\begin{effectivemodel} \label{model:GCC-effective-model}
{\bf Gaussian collective coordinate (GCC) functional form.}
\begin{align}
    \dot{z}_\sigma = &\left(\left\langle \iu \omega_i \right\rangle_{i\in P_\sigma} -\sqrt{-2v_\sigma\ln(|z_\sigma|)} \right)z_\sigma \nonumber \\
    &+ \sum_{\sigma'=1}^{C} |P_{\sigma'}|K_{\sigma \sigma'} \left( z_{\sigma'} - z_{\sigma'}^*z_\sigma^2 |z_\sigma|^2 \right),
    \label{eq:closed_CG_with_gaussian_cc}
\end{align}
is the first class of closed reduced-order system of equations considered here. It was derived by substituting the closure approximations \eqref{eq:A-closure-GCC} and \eqref{eq:F-closure-GCC} into Eq. \eqref{eq:reduced_eqn_terms_gathered}. We refer to the functional form on the right-hand side of Eq. \eqref{eq:closed_CG_with_gaussian_cc} as the Gaussian collective coordinate (GCC) form.
\end{effectivemodel}

On the other hand, it is reasonable to suppose that the equations \eqref{eq:skardal_restrepo_CG_model} derived by Ott and Antonsen, proven to be valid in the $N\to \infty$ limit, are still approximately valid for small $N$, or at least are worthy of consideration as a candidate model class.

\begin{effectivemodel} \label{model:COA-effective-model}
{\bf Cauchy Ott-Antonsen (COA) functional form.}
\begin{align}
    \dot{z}_\sigma = & \left( \left\langle \iu \omega_i \right\rangle_{i\in P_\sigma} -\delta_\sigma \right)z_\sigma \nonumber \\
    & + \sum_{\sigma'=1}^{C} |P_{\sigma'}|K_{\sigma \sigma'} \left( z_{\sigma'} - z_{\sigma'}^*z_\sigma^2 \right).
    \label{eq:COA}
\end{align}
is the second class of closed reduced-order system of equations considered here. It is derived by applying the Ott-Antonsen ansatz to the infinite-$N$ limit of Eq. \eqref{eq:kuramoto_modular} under the assumption of Cauchy-distributed natural frequencies within each module. We refer to the functional form on the right-hand side of Eq. \eqref{eq:COA} as the Cauchy Ott-Antonsen (COA) form.
\end{effectivemodel}

It is in fact possible to formally derive Eq. (\ref{eq:COA}) by applying the formulae (\ref{eq:general_cc_closure_terms}) under the assumption that natural frequencies in each cluster $\sigma$ are Cauchy distributed with location $\Omega_\sigma$ and scale $\delta_\sigma$. In this case we have $\chi_\sigma(t) = \exp(-\delta_\sigma |t|)$, so
\begin{align}
    \langle\iu \omega_i x_i \rangle &= - \delta_\sigma z_\sigma\\
    \langle x_i^2 \rangle &= 0.
\end{align}
{Note, however, that strictly speaking, the collective coordinate approximation (\ref{eq:collective_coord_ansatz}) is not consistent in this case because it models the phases of oscillators with large natural frequencies to be a large constant greater than the mean phase, and implies that their instantaneous frequency is the mean frequency, but this is not the case.}

Since in general we consider small numbers of oscillators, it is not necessarily true that the set of natural frequencies within any given cluster is representative of a Gaussian- or Cauchy-distributed population. To circumvent this, we approximate the characteristic function $\chi_\sigma$ as a Taylor series, and show that {when $\alpha$ is small, $\chi_\sigma(\alpha)$ is determined (to a good approximation) by} the first and second central moments of the empirical distribution of natural frequencies within a cluster. We have
\begin{align}
    \chi_\sigma(\alpha) &= \left\langle\exp(\iu \alpha \tilde{\omega}_i)\right\rangle_{i\in P_\sigma}\\
    &= \left\langle\sum_{k=0}^{\infty} \frac{(\iu \alpha \tilde{\omega}_i)^k}{k!} \right\rangle_{i\in P_\sigma}\\
    &= \sum_{k=0}^{\infty} \frac{(\iu \alpha)^k}{k!} \left\langle\tilde{\omega}_i^k\right\rangle_{i\in P_\sigma}\\
    & = 1 + \iu \alpha \langle \tilde{\omega}_i \rangle_{i\in P_\sigma} - \frac{\alpha^2}{2} \langle \tilde{\omega}_i^2 \rangle_{i\in P_\sigma} + \mathcal{O}(\alpha^3)
\end{align}
Note that the relevant values of $\alpha$ are necessarily such that the residual phases $\alpha \tilde{\omega}_i$ are not too big, and so are inversely related with the largest values of $\tilde{\omega}_i$. Finally we must impose the consistency condition (\ref{eq:consistency-relation-for-alpha}). Assuming that terms of $\mathcal{O}(\alpha^3)$ are negligible and that $\langle \tilde{\omega}_i\rangle = 0$, we can solve the consistency condition $\chi_\sigma(\alpha_\sigma) = |z_\sigma|$ for $\alpha_\sigma$ to get
\begin{equation}
    \alpha_\sigma \approx \sqrt{\frac{2(1-|z_\sigma|)}{\langle \tilde{\omega}_i^2\rangle_{i\in P_\sigma}}}.
\end{equation}
Using this approximate solution to estimate the closure terms in Eq. \eqref{eq:general_cc_closure_terms}, we obtain
\begin{align}
    \dot{z}_\sigma = & \left( \left\langle \iu \omega_i \right\rangle_{i\in P_\sigma} - \frac{1}{|z_\sigma|}\sqrt{2(1-|z_\sigma|)\langle \tilde{\omega}_i^2 \rangle }\right)z_\sigma \nonumber \\
    & + \sum_{\sigma'=1}^{C} |P_{\sigma'}|K_{\sigma \sigma'} \left( z_{\sigma'} - z_{\sigma'}^*z_\sigma^2 \frac{4|z_\sigma| - 3}{|z_\sigma|^2} \right),
    \label{eq:closed_CG_with_general_cc_taylor_expanded}
\end{align}
which approximates Eq. \eqref{eq:closed_CG_with_gaussian_cc} when $\{|z_\sigma|\}$ are near 1.

\subsection{Data-driven Inference of Model Parameters} \label{sec:inference}

The derivations above make several simplifying assumptions and useful approximations, some of which will not be satisfied in practice. To find a coarse-grained description of a given finite-dimensional system of coupled oscillators, we take a hybrid approach, using the above derivations as inductive biases, and using data to select coefficient values.

We focus especially on the coarse-grained systems in Eq. \eqref{eq:closed_CG_with_gaussian_cc}, i.e., GCC, and Eq. \eqref{eq:COA}, i.e., COA. For each coarse-grained system, the right-hand side is a combination of particular functions of $\{z_\sigma\}$, each with a coefficient that depends somehow on the distribution of $\{\omega_i\}$ and the coupling matrix $K$. In the absence of the $N\to \infty$ limit, however, it is not clear that simple summary statistics will give a model that accurately reflects the coarse-grained dynamics of the finite-$N$ system.

We now describe a procedure for inferring the parameters of a coarse-grained model, either of the form in Eq. \eqref{eq:closed_CG_with_gaussian_cc} or Eq. \eqref{eq:COA} assuming that the coarse-graining partition $\Pi$ is known. 
To illustrate what we mean explicitly, we restrict our attention momentarily to Eq. \eqref{eq:COA}. Given a solution $\{\theta_i(t)\}$ of the system \eqref{eq:kuramoto_modular} in a time-series form, we can obtain a coarse-grained time series $\{z_\sigma(t)\}$ according to Eq. \eqref{eq:local_order_params}. Our hypothesis is that these coarse-grained variables evolve according to an equation of the form Eq. \eqref{eq:COA}. If the hypothesis holds, then it should be possible to recover the appropriate coefficients, which we can interpret as effective natural frequencies and coupling parameters, using a \emph{least-squares regression} based on the  coarse-grained data $\{z_\sigma(t)\}$.  Observe first that the right-hand side of \eqref{eq:COA} is a linear combination of terms that can be measured directly from data. To clarify this, we re-write Eq. \eqref{eq:COA} as
\begin{equation}
\dot{z}_\sigma = \tilde{\omega}_\sigma z_\sigma +\sum_{\sigma' = 1}^C B_{\sigma \sigma'} \left(z_{\sigma'} - z_{\sigma'}^* z_\sigma^2\right),
\label{eq:skardal_restrepo_CG_compressed}
\end{equation}

Let $\boldsymbol{z}_\sigma=(z_\sigma(t_1), \dots, z_\sigma (t_n))^T$ and $\dot{\boldsymbol{z}}_\sigma=(\dot{z}_\sigma(t_1), \dots, \dot{z}_\sigma (t_n))^T$ denote the time-series of observations of the coarse-grained variable $z_\sigma$ and its derivative, respectively.  Then the parameters $\boldsymbol{p}_\sigma = (\tilde{\omega}_\sigma, B_{\sigma 1},\dots, B_{\sigma C})^T$ can be inferred by solving the following least-squares optimization \cite{Brunton2015, lokhov2018online}
\begin{align}
    \argmin_{\boldsymbol{p}_\sigma} &\Vert \dot{\boldsymbol{z}}_\sigma - \boldsymbol{G}^\sigma \boldsymbol{p}_\sigma \Vert^2 \label{eq:least-squares} \\
    \text{s.t.} \quad & \begin{cases}
        \Im(\tilde{\omega_\sigma}) \ge 0\\
        \Im(B_{\sigma \sigma'}) = 0 \label{eq:least_squares_constraints}
    \end{cases}
\end{align}
where the $n\times C+1$ matrix $\boldsymbol{G}^\sigma$ contains the values of the terms of which the entries of $\boldsymbol{p}_\sigma$ are coefficients, and $\Im(\cdot)$ denotes the imaginary part of the argument. Explicitly, we can write the set of basis functions $\boldsymbol{G}^\sigma$  as
\begin{equation}
    \boldsymbol{G}^\sigma = \left[
    \begin{array}{c|c|c|c}
          \boldsymbol{z}_\sigma & \boldsymbol{z}_1 - \boldsymbol{z}_1^* \boldsymbol{z}_\sigma^2 & \dots & \boldsymbol{z}_C - \boldsymbol{z}_C^* \boldsymbol{z}_\sigma^2\\
    \end{array}
    \right]
\end{equation}
where complex conjugate, product, and squaring of the vectors $\boldsymbol{z}_\sigma$ are understood to be taken element-wise. Minimizing the squared residual \eqref{eq:least-squares} for each $\sigma=1,\dots,C$ gives a complete set of parameters for the COA model in Eq. \eqref{eq:skardal_restrepo_CG_compressed}. The constraints in Eq. (\ref{eq:least_squares_constraints}) ensure that the resulting model leaves the $C$-fold product of the unit disk invariant, i.e. that $|z_\sigma(t)|\le 1 \: \forall \sigma \implies |z_\sigma(s)| \le 1 \: \forall \sigma$ for all $s>t$. This property is desirable since $\{z_\sigma\}$ are averages of unit complex numbers, and so should always lie inside the unit disk.

We can likewise perform the same analysis using the inductive bias given by GCC in Eq. \eqref{eq:closed_CG_with_gaussian_cc}, by setting the set of basis functions as follows:
\begin{equation}
    \boldsymbol{G}^\sigma = \left[
    \begin{array}{c}
          \boldsymbol{z}_\sigma \\
          \sqrt{-2\ln(|z_\sigma|)} \\
          \boldsymbol{z}_1 - \boldsymbol{z}_1^* \boldsymbol{z}_\sigma^2 |\boldsymbol{z}_\sigma^2|\\
          \vdots \\
          \boldsymbol{z}_C - \boldsymbol{z}_C^* \boldsymbol{z}_\sigma^2|\boldsymbol{z}_\sigma^2|\\
    \end{array}
    \right]^T
\end{equation}

In accordance with previous results \cite{ott2008low,Skardal2012}, we expect to see a satisfactory fit if $N$ (the number of oscillators) is very large and the coupling is as described in Eq. \eqref{eq:kuramoto_modular}. The numerical procedure described above lets us assess the suitability of GCC in Eq. \eqref{eq:closed_CG_with_gaussian_cc} and COA in Eq. \eqref{eq:skardal_restrepo_CG_compressed} as coarse-grained models for finite $N$.

\section{Results} \label{sec:results}

We now evaluate the performance of the above-described coarse-grained models by applying them to a range of Kuramoto oscillator systems for which we have fine-grained trajectories. We examine their performance both with theoretically-derived coefficient values (as in Eqs. (\ref{eq:closed_CG_with_gaussian_cc}) and (\ref{eq:COA})) as well as those optimized to fit observations. Further, we demonstrate the necessity of both Conditions \ref{ass:modularity-of-coupling-matrix} and \ref{ass:phase-cohesive-clusters} by showing that coarse-graining in a way that satisfies only one or the other Condition gives a worse model than coarse-graining in a way that satisfies both.

\subsection{Data Generation}

\begin{figure}
    \centering
    \includegraphics[width=\columnwidth]{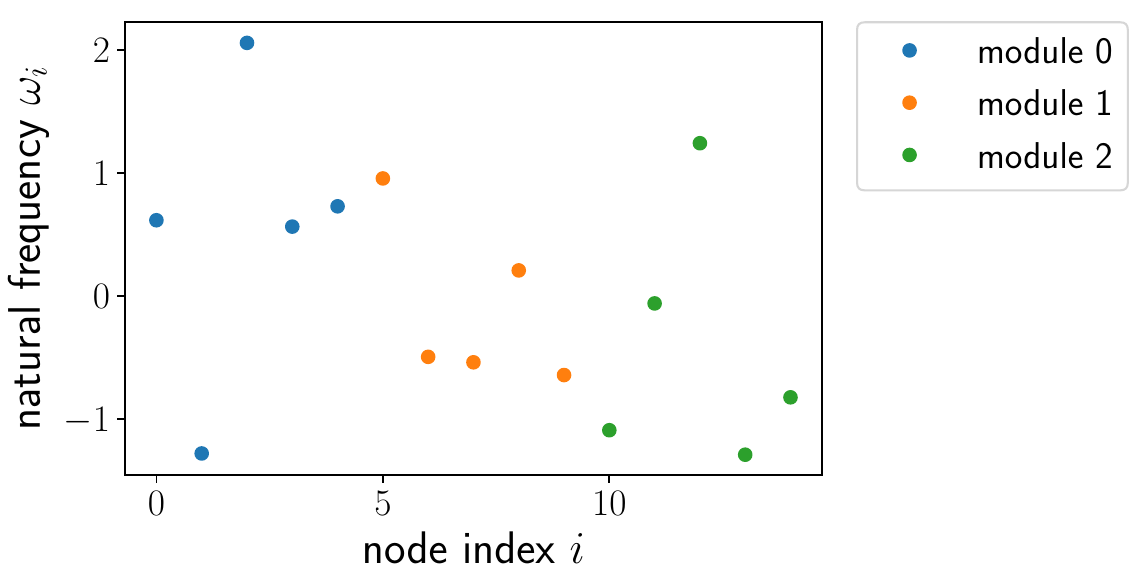}
    \caption{Natural frequencies used in numerical experiments}
    \label{fig:natural_frequencies_scatterplot}
    \vspace{5ex}
\end{figure}

We perform numerical experiments for a system of $N=15$ Kuramoto oscillators, i.e. Eq. (\ref{eq:kuramoto}), organized into three modules of five nodes each, and we only consider coupling matrices that respect this organization. This means that we know the \emph{structural partition} $\Pi^\text{struct}$ a priori; it is $\{\{1,\dots,5\}, \{6,\dots,10\}, \{11,\dots,15\}\}$. The coupling between nodes is such that each pair of nodes in the same module is coupled with strength $K_{ij} = K_\text{in} / N$ and pairs of nodes in different modules are coupled with strength $K_{ij} = K_\text{out}/ N$. For each oscillator we sample a natural frequency $\omega_i$ from a standard normal distribution (i.i.d.), and fix them through all experiments, and their values are shown in Fig. \ref{fig:natural_frequencies_scatterplot}. The free parameters in our experiments are $K_\text{in} \in [0, 10]$ and $K_\text{out} \in [0, 2.5]$; for each parameter we sampled 50 values evenly spaced in the respective interval (inclusive of endpoints). For each pair of $(K_\text{in}, K_\text{out})$ values, we generate a trajectory of length $T=1000$ with a time resolution of $\Delta t = 0.01$, starting from uniformly random initial phases.

\subsection{Coarse-Grained Model Selection}

\begin{figure}
    \centering
    \includegraphics[width=\columnwidth]{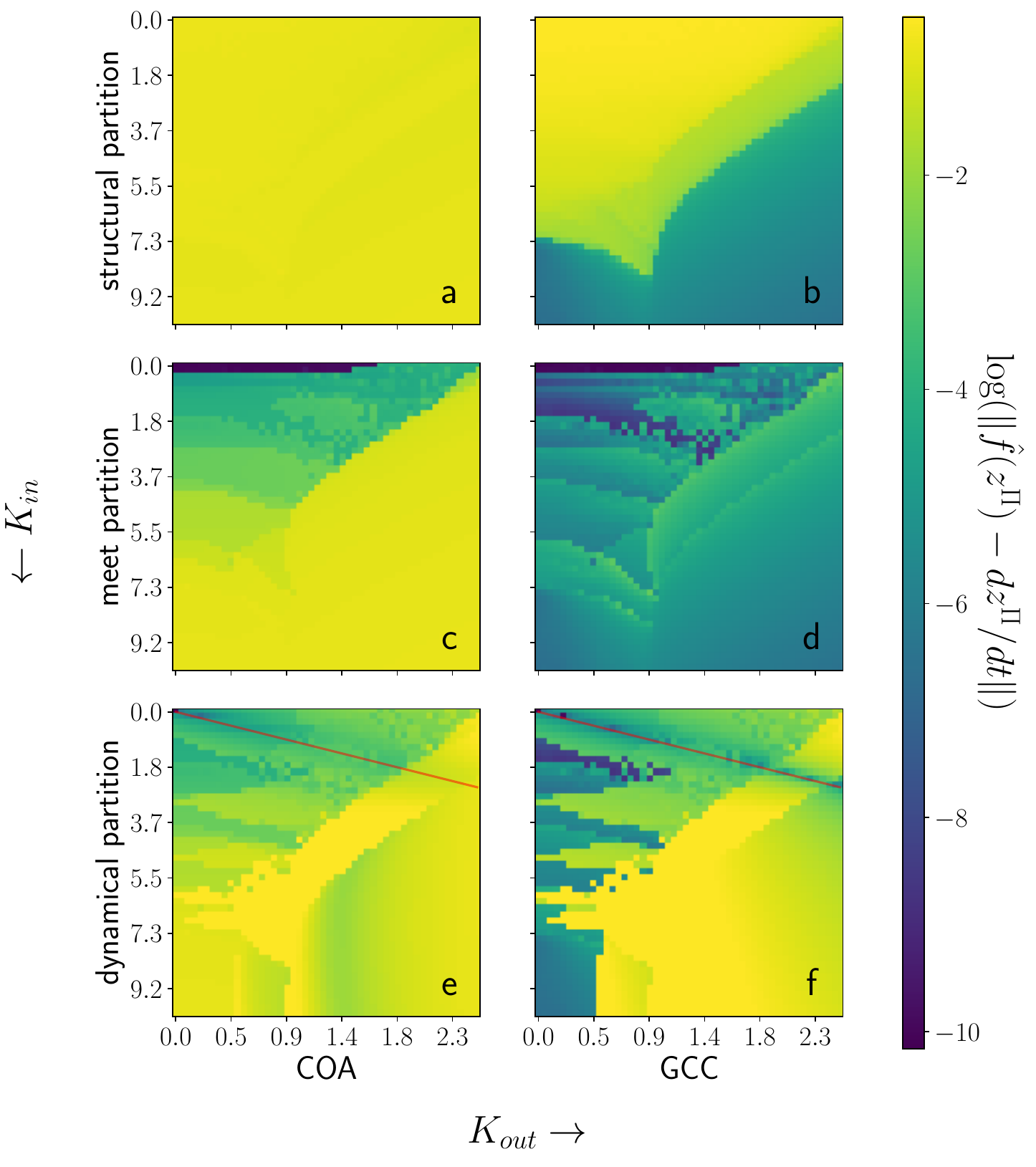} \vspace{-5ex}
    \caption{Coarse-grained model quality as a function of coupling parameters for theoretical coefficient values. }
    \label{fig:GCC_vs_COA_theory_heatmaps}
\end{figure}

\begin{figure}
    \centering
    \includegraphics[width=\columnwidth]{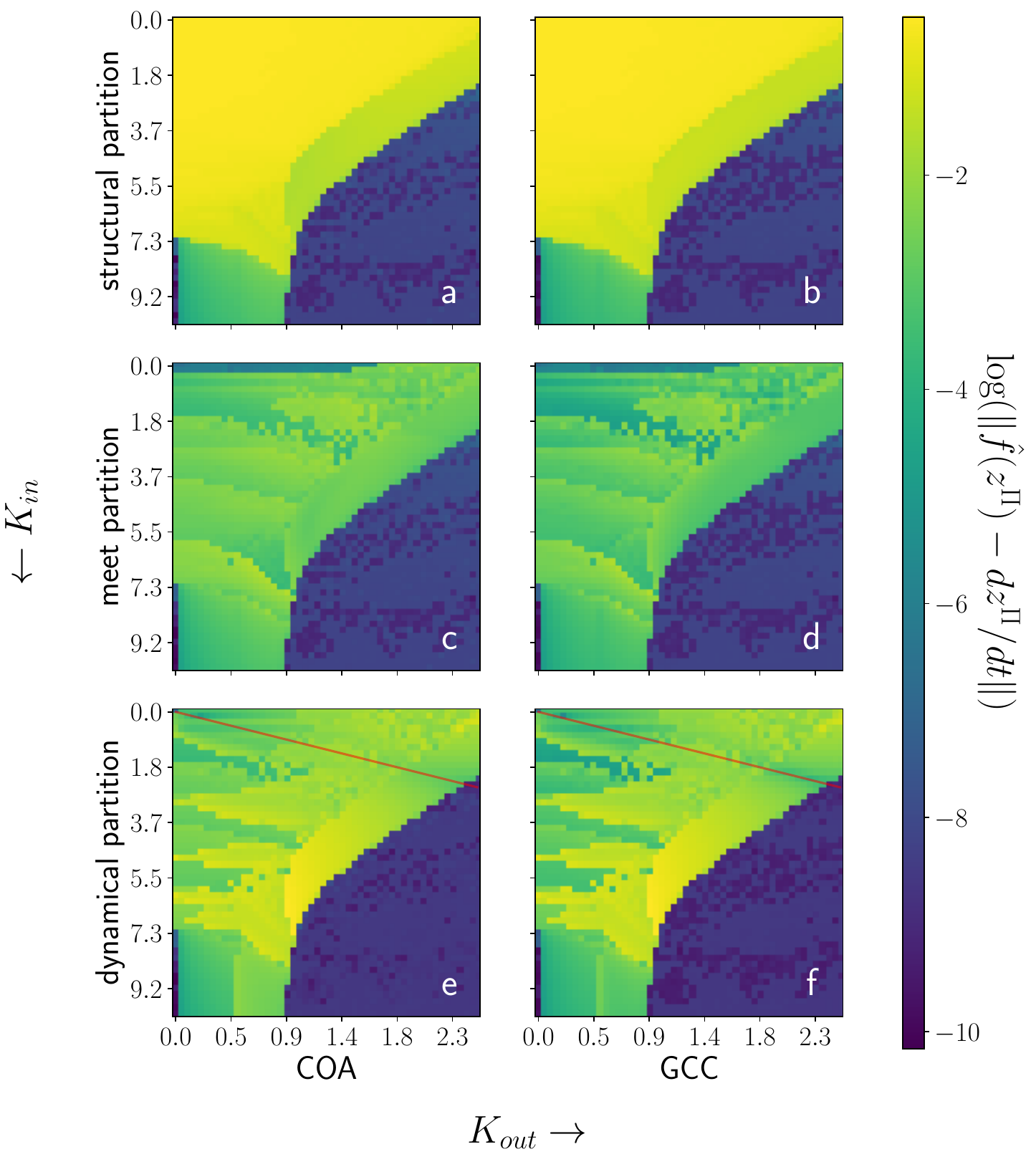} \vspace{-5ex}
    \caption{As in Fig. \ref{fig:GCC_vs_COA_theory_heatmaps} but for coefficient values optimized to fit the data.}
    \label{fig:GCC_vs_COA_inference_heatmaps}
\end{figure}

For each trajectory, we derive a coarse-grained model in each of twelve ($3\cdot 2 \cdot 2$) different ways, corresponding to three different choices:
\begin{itemize}
    \item Partition: structural (Figs. \ref{fig:GCC_vs_COA_theory_heatmaps} and \ref{fig:GCC_vs_COA_inference_heatmaps}, a-b) vs. dynamical (Figs. \ref{fig:GCC_vs_COA_theory_heatmaps} and \ref{fig:GCC_vs_COA_inference_heatmaps}, e-f) vs. meet (Figs. \ref{fig:GCC_vs_COA_theory_heatmaps} and \ref{fig:GCC_vs_COA_inference_heatmaps}, c-d)
    \item Model class: Cauchy-Ott-Antonsen (COA) (Figs. \ref{fig:GCC_vs_COA_theory_heatmaps} and \ref{fig:GCC_vs_COA_inference_heatmaps}, a, c, e)  vs. Gaussian-collective coordinate (GCC) (Figs. \ref{fig:GCC_vs_COA_theory_heatmaps} and \ref{fig:GCC_vs_COA_inference_heatmaps}, b, d, f)
    \item Method for finding coefficients: theory (Fig. \ref{fig:GCC_vs_COA_theory_heatmaps}) vs. inference (Fig. \ref{fig:GCC_vs_COA_inference_heatmaps})
\end{itemize}
In what follows we clarify the meaning of each of these choices and examine their effect on the quality of coarse-grained model that results.

\begin{figure}
    \centering
    \includegraphics[width=\columnwidth]{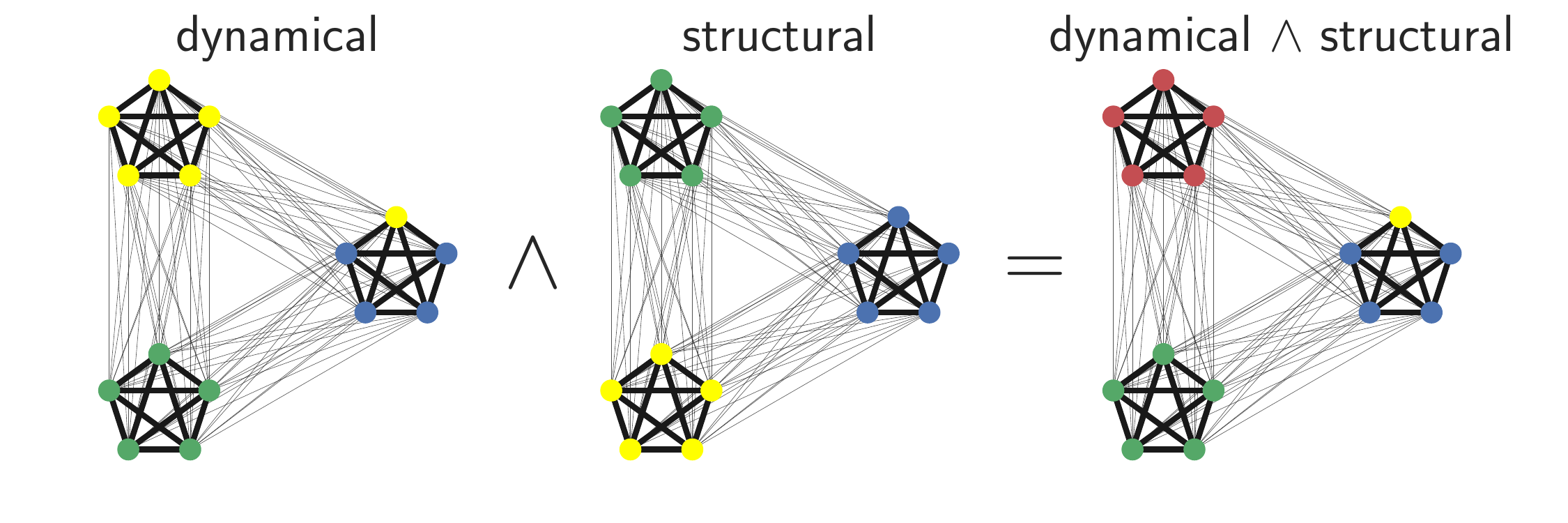} \vspace{-5ex}
    \caption{An example of the meet of the structural and dynamical partitions. In the structural partition, nodes are colored according to the partition that determines the coupling strengths (line thicknesses) between nodes. In the dynamical partition, nodes are colored according to their membership in phase-cohesive groups. The meet (right) is the coarsest partition that refines both the structural and dynamical partitions.}
    \label{fig:partition_meet_example}
    \vspace{5ex}
\end{figure}

\textbf{Partition:} For each trajectory $\{\theta_i(t)\}$, we must choose a partition to coarse-grain the data. According to Condition \ref{ass:modularity-of-coupling-matrix}, the partition should respect the symmetry of the coupling matrix, i.e. it should be a refinement of $\Pi^\text{struct}$, and according to Condition \ref{ass:phase-cohesive-clusters} it should be such that all oscillators within a partition element are phase locked, i.e. it should be a refinement of $\Pi^\text{dyn}$. Imposing both conditions means that the partition should refine both $\Pi^\text{struct}$ and $\Pi^\text{dyn}$, i.e. it should refine $\Pi^\text{meet} \coloneqq \Pi^\text{struct} \wedge \Pi^\text{dyn}$ (see Fig. \ref{fig:partition_meet_example}). In order to illuminate the importance of each Condition while obtaining the coarsest possible model, we evaluate the performance models coarse-grained according to each of $\Pi^\text{struct}, \Pi^\text{dyn}$, and $\Pi^\text{meet}$.

In practice, it is prohibitively expensive to enumerate all partitions of $N$ nodes to find $\Pi^\text{dyn}$ that satisfies the condition defined by Eq. \eqref{eq:dynamical_equivalence_relation}, so we use a heuristic.
Given a trajectory $\{\theta_i(t)\}$, we compute a long-term average frequency $\overline{\omega}_i = (\theta_i(t_0 + T) - \theta_i(t_0))/(T-t_0)$, and cluster nodes initially by their value of $\overline{\omega}$. We can then check if the resulting clusters are in fact phase cohesive, and if they are not, perform a finer clustering of $\overline{\omega}$ and repeat until we have phase cohesive clusters. Other approaches to identifying phase clusters have been considered and could be used in place of the method described above \cite{Bialonski2006,Groth2011,Allefeld2007}. Constructing $\Pi_\text{meet}$ from $\Pi_\text{struct}$ and $\Pi_\text{dyn}$ is straightforward; it is simply the set of all nonempty pairwise intersections of partition elements from $\Pi_\text{struct}$ and $\Pi_\text{dyn}$.

\textbf{Model class:} Given a partition $\Pi$, we coarse-grain the data to obtain $\{z_\sigma(t) | P_\sigma \in \Pi\}$. We now choose what class of model to find, COA or GCC. The COA model is defined by Eq. (\ref{eq:COA}), and the GCC model is defined by Eq. (\ref{eq:closed_CG_with_gaussian_cc}).

\textbf{Method for finding coefficients:} Finally, we have the choice of using coefficient values predicted from theory (that includes several possibly-violated assumptions and approximations), or inferring them from data as described in Sec. \ref{sec:inference}. Note that the theoretical values present in the COA model depend on parameters $(\Omega_\sigma, \delta_\sigma)$ of the Cauchy distributions from which natural frequencies are drawn, which do not actually exist in our case. To circumvent this difficulty we perform a maximum-likelihood inference of these parameters from the data $\{\omega_i | i\in P_\sigma \}$ assuming that they were drawn from a Cauchy distribution (as implemented by \texttt{scipy.stats.cauchy.fit} \cite{Virtanen2020}).

\begin{figure}
    \centering
    \includegraphics[width=\columnwidth]{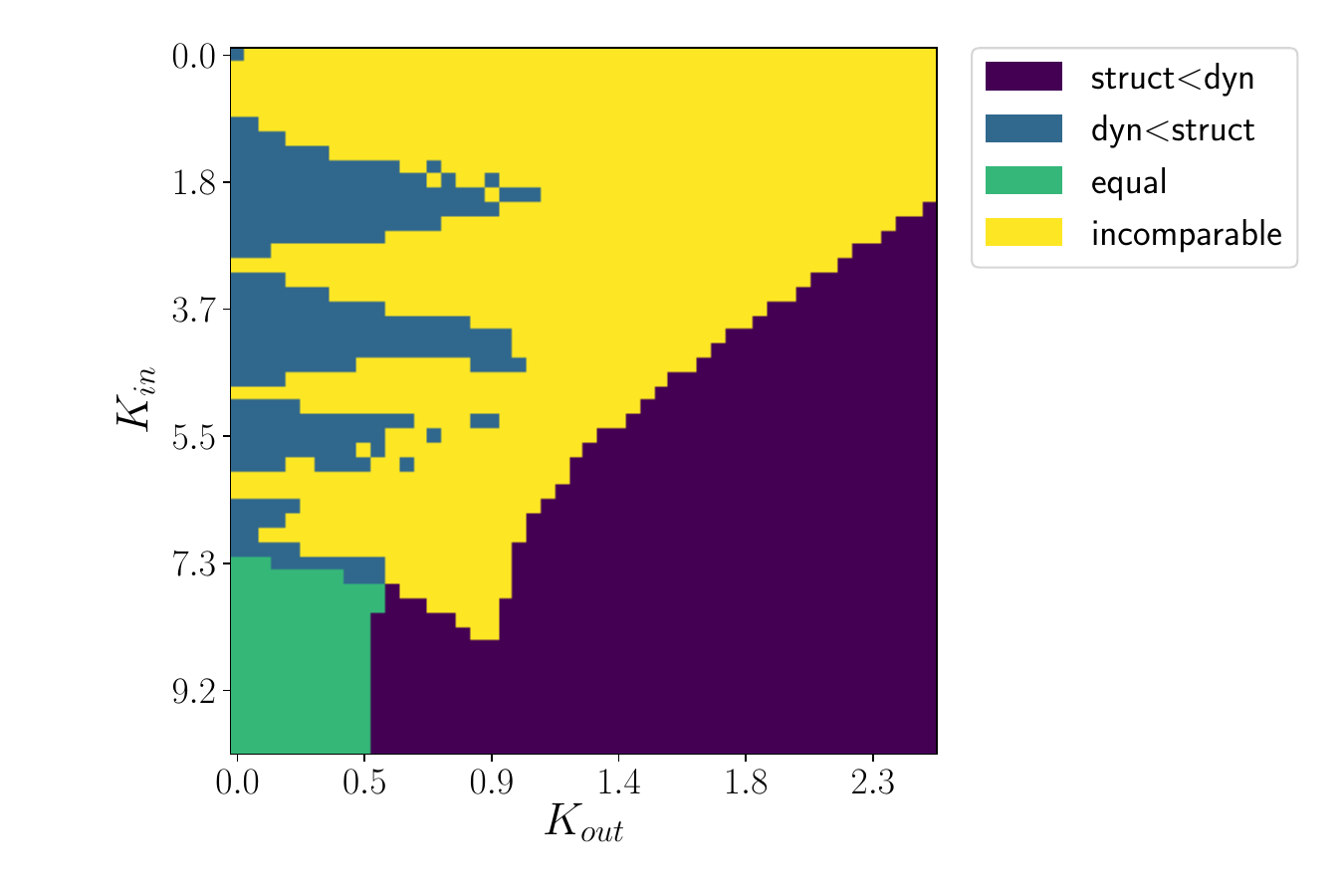} \vspace{-5ex}
    \caption{Parameter space $(K_\text{in},K_\text{out})$ colored by how the dynamical partition compares to the structural partition. The relation $P<Q$ means that partition $P$ is a refinement of partition $Q$, and incomparable means that neither $P<Q$ nor $Q<P$. See Fig. \ref{fig:partition_comparability_cases} for examples of each of these cases.}
    \label{fig:partition_comparison_heatmap}
\end{figure}

Overviews of coarse-grained model quality as a function of coupling parameters $(K_\text{in}, K_\text{out})$ are depicted in Figs. \ref{fig:GCC_vs_COA_theory_heatmaps} and \ref{fig:GCC_vs_COA_inference_heatmaps}. The metric used is root-mean-squared error in time-derivative, i.e. $\Vert \hat{f}(z^\Pi) - \du z^{\Pi} / \du t\Vert$, where $\hat{f}$ is the coarse-grained model in question and $\Vert \cdot \Vert$ is the root-mean-square with respect to time $t$ and partition element index $\sigma$. Heatmap colors are arranged on a logarithmic scale. Compare to Fig. \ref{fig:partition_comparison_heatmap} to relate coarse-grained model performance to characteristics of the dynamical partition relative to the structural partition.

\begin{figure}
    \centering
    \includegraphics[width=\columnwidth]{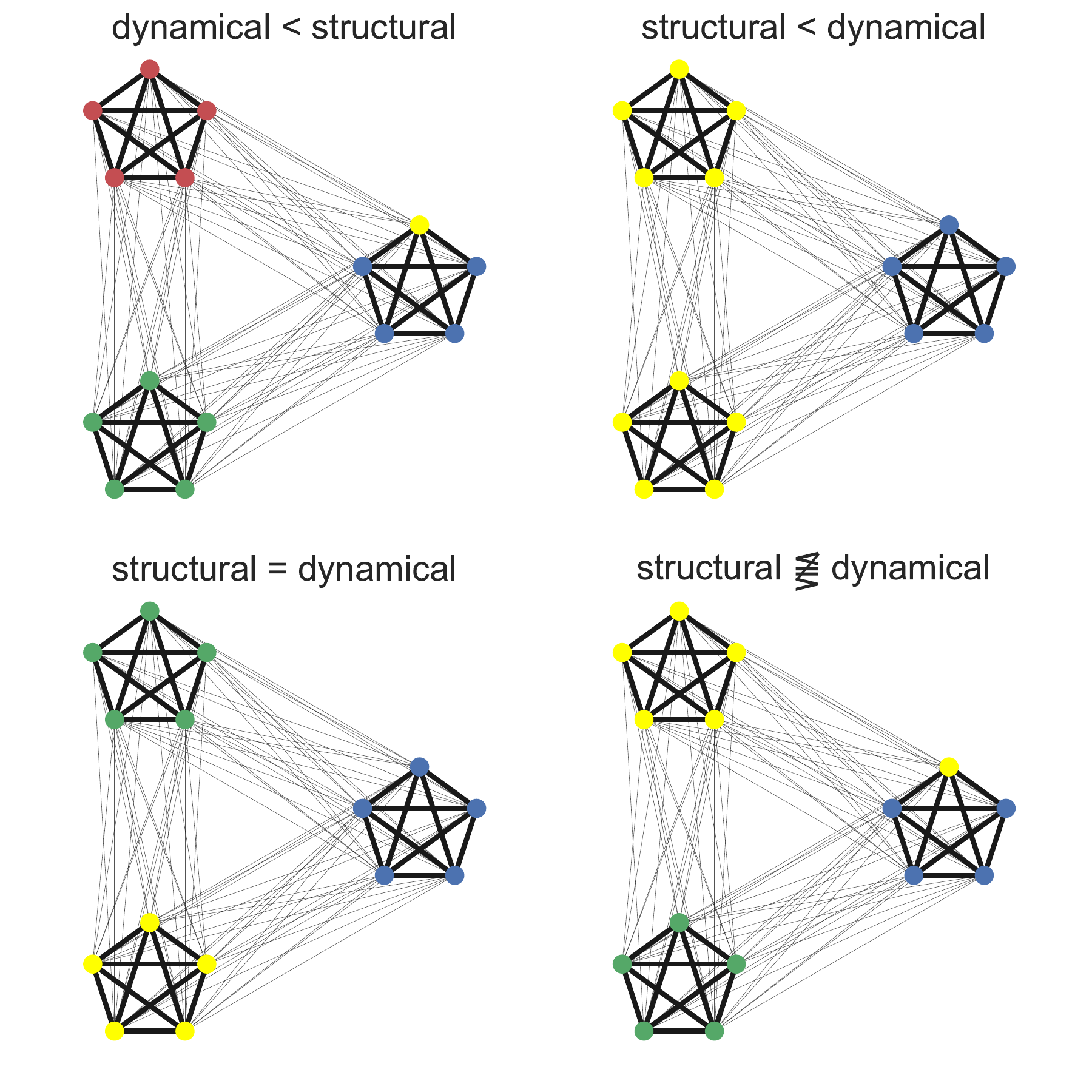} \vspace{-5ex}
    \caption{Depiction of the possible relationships between two partitions. Here the structural partition is the partition into three groups of five nodes indicated by their position and edge thickness, and color indicates the dynamical partition, i.e. nodes of the same color are phase-locked with each other.}
    \label{fig:partition_comparability_cases}
    \vspace{5ex}
\end{figure}

A more systematic set of comparisons illuminating the impact of the three choices (partition, model terms, and method) is given in Figures \ref{fig:log-rel-loss-theory_v_inference}-- \ref{fig:log-loss-meet_v_dynamical}.

\subsection{Numerical Results}

We now present an overview of the dynamics we simulated, and the performance of each of the twelve coarse-grained models described above.

While $\Pi^\text{struct}$ is known a priori and fixed for all experiments, $\Pi^\text{dyn}$ changes depending on the values of $K_\text{in}$ and $K_\text{out}$. Figure \ref{fig:partition_comparison_heatmap} gives an overview of how $\Pi^\text{dyn}$ compares to $\Pi^\text{struct}$ as a function of $K_\text{in}$ and $K_\text{out}$. There are four possibilities: either $\Pi^\text{struct} < \Pi^\text{dyn}$, $\Pi^\text{dyn}<\Pi^\text{struct}$, $\Pi^\text{struct} = \Pi^\text{dyn}$, or the two partitions are incomparable (i.e. neither one is a refinement of the other, nor are they equal). As one might expect, the region of parameter space where the two partitions are equal lies where $K_\text{in}$ is large and $K_\text{out}$ is small. When both $K_\text{in}$ and $K_\text{out}$ are large, global synchronization occurs and thus $\Pi^\text{struct}<\Pi^\text{dyn}$ holds. For moderate values of $K_\text{in}$ and small $K_\text{out}$, phase-locked clusters remain restricted to structural clusters and $\Pi^\text{dyn}<\Pi^\text{struct}$, while for the remaining parameter space the two partitions are incomparable.

\textbf{Theoretically determined coefficients:} In Fig. \ref{fig:GCC_vs_COA_theory_heatmaps}, we give an overview of coarse-grained model quality using theoretically-derived coefficient values. We can see that the same subsets of parameter space highlighted in Fig. \ref{fig:partition_comparison_heatmap} appear in panels b, d, and f of Fig. \ref{fig:GCC_vs_COA_theory_heatmaps}, corresponding to the GCC ansatz. Note that applying the GCC ansatz to the meet partition (d) obtains reasonably good performance across the entire parameter space, while under the structural partition (b), good performance is only attained when the structural modules are in fact internally synchronized. Conversely, under the dynamical partition (f), the GCC model performs well mainly in regions of parameter space where the dynamical partition refines or is equal to the structural one (cf. Fig. \ref{fig:partition_comparison_heatmap}), since in these cases the dynamical partition equals the meet partition. Notice also that when $K_\text{in} \approx K_\text{out}$ (red line, e-f), the structural partition is effectively the entire network as a single module, and in this region we also see somewhat better performance of the GCC model. The COA ansatz attains reasonable performance for small values of $K_\text{in}$ under the meet partition - in this case many dynamical modules are singletons, and are therefore trivial to coarse-grain; otherwise, the COA ansatz does not perform very well, indicating that the assumptions underlying its derivation do not hold for the system at hand.

\textbf{Coefficients optimized to fit the data:} In Fig. \ref{fig:GCC_vs_COA_inference_heatmaps} we give an overview of coarse-grained model quality using coefficient values optimized to fit the data. The most notable contrast with Fig. \ref{fig:GCC_vs_COA_theory_heatmaps} is that the left and right columns (COA vs. GCC) are much more similar to each other in Fig. \ref{fig:GCC_vs_COA_inference_heatmaps} than in Fig. \ref{fig:GCC_vs_COA_theory_heatmaps}. This indicates that given the possibility of tuning coefficient values, the functional form of the COA ansatz is nearly equivalently expressive to the functional form of the GCC ansatz, relative to the dynamics at hand. Again, note that coarse-graining according to the dynamical (resp. structural) partition attains good performance in regions where it refines the structural (resp. dynamical) partition and hence is equal to the meet partition. However, all partitions are able to achieve good performance in the large region of global synchrony, owing to the simplicity of the resulting dynamics.

\begin{figure}
    \centering
    \includegraphics[width=\columnwidth]{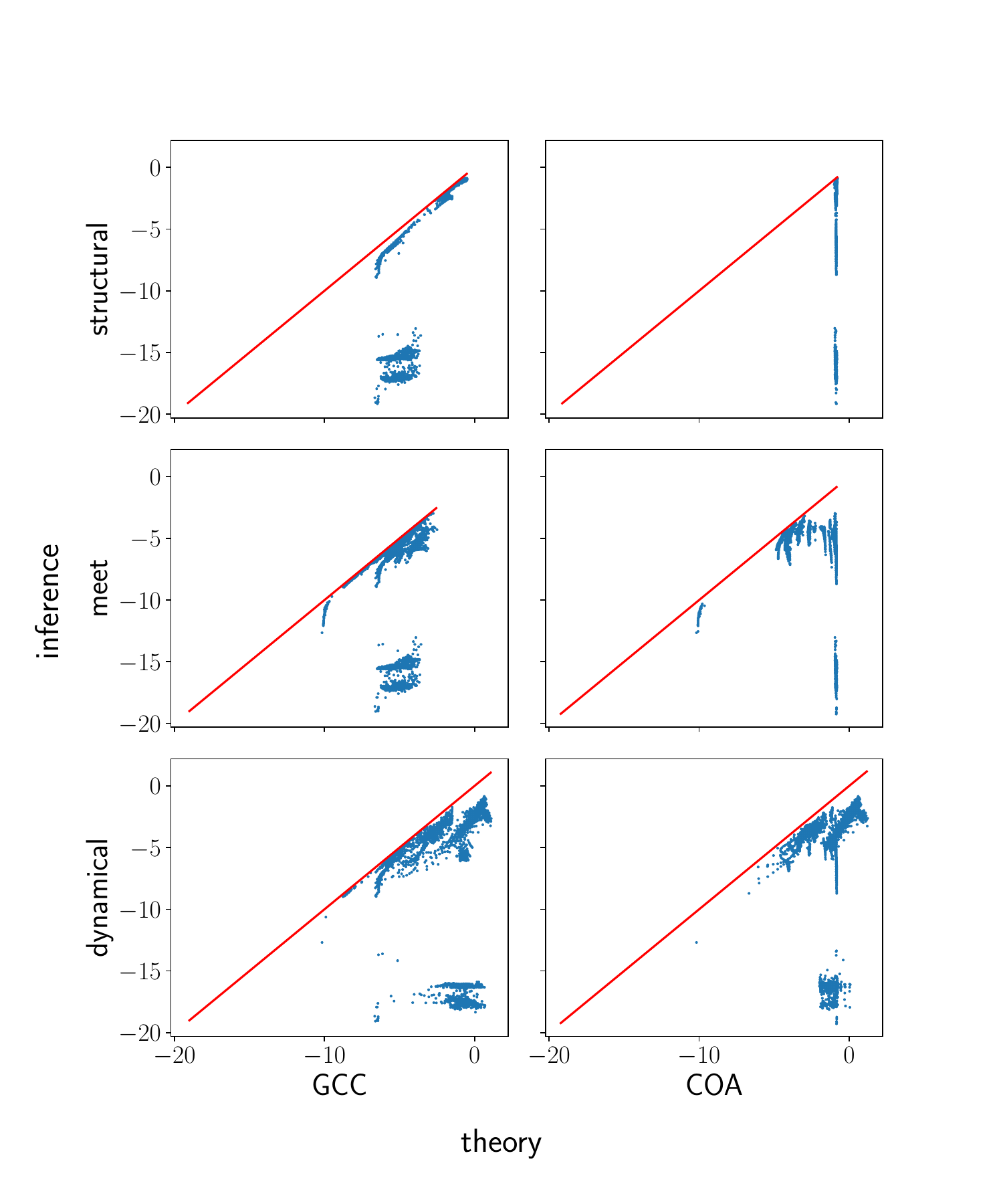} \vspace{-5ex}
    \caption{
    Log-loss of the coarse-grained model obtained by inferring coefficient values vs. using theoretical values. Each point corresponds to a single instance of the coarse-graining problem. In many cases (under the meet partition) the GCC theory performs nearly as well as GCC inference, indicating that the assumptions going into the GCC theory were well-supported. Note that COA theory with the structural partition never does particularly well.}
    \label{fig:log-rel-loss-theory_v_inference}
    \vspace{5ex}
\end{figure}

\begin{figure}
    \centering
    \includegraphics[width=\columnwidth]{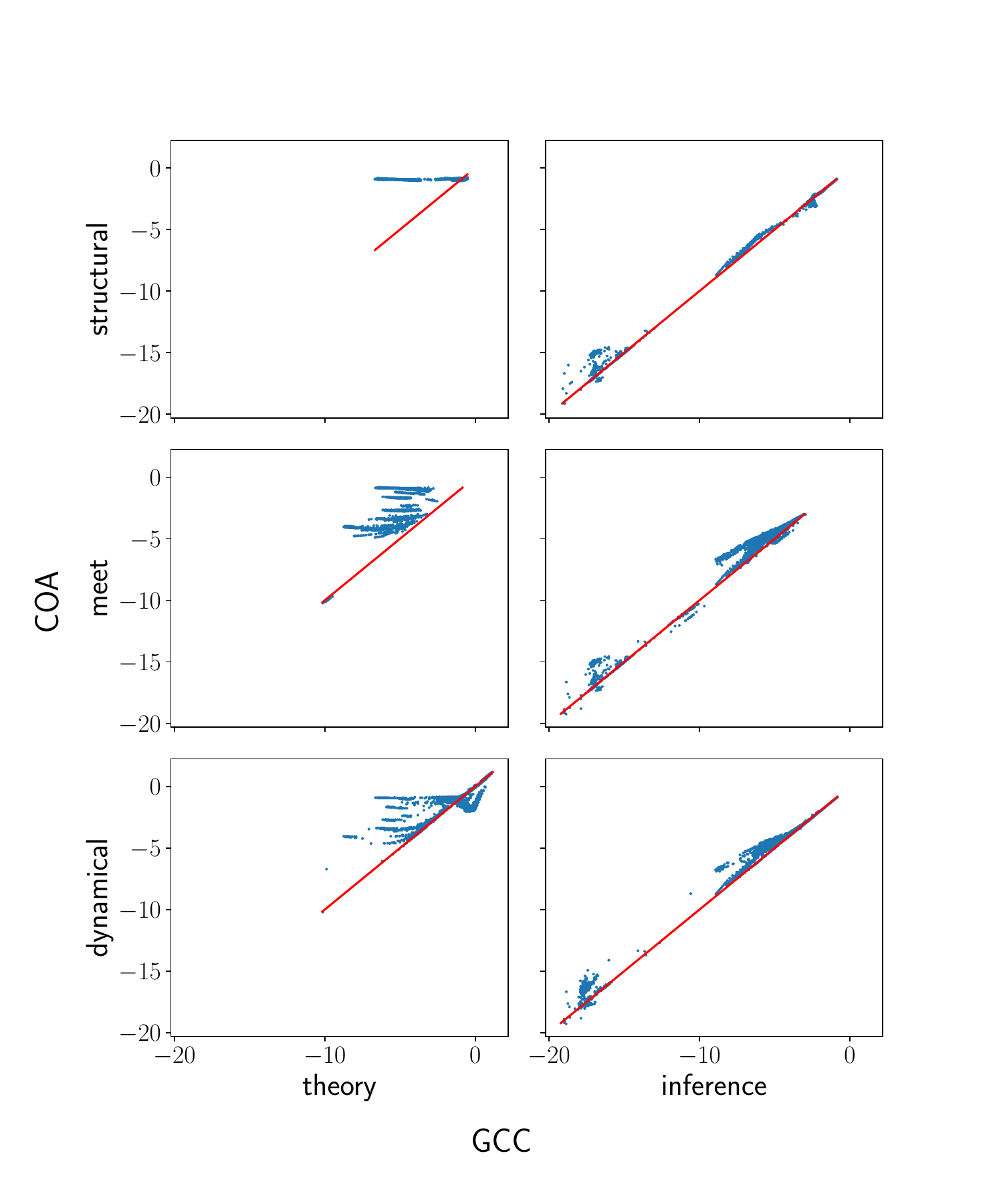} \vspace{-5ex}
    \caption{As in Fig. \ref{fig:log-rel-loss-theory_v_inference}, organized to show the effect of using model terms predicted from GCC instead of COA. Interestingly, when given the freedom to adapt coefficient values to observed data, COA and GCC models perform comparably. This indicates that relative to the behaviors considered in these numerical experiments, the COA and GCC function spaces are somehow comparably expressive.}
    \label{fig:log-rel-loss-COA_v_GCC}
    \vspace{5ex}
\end{figure}

\begin{figure}
    \centering
    \includegraphics[width=\columnwidth]{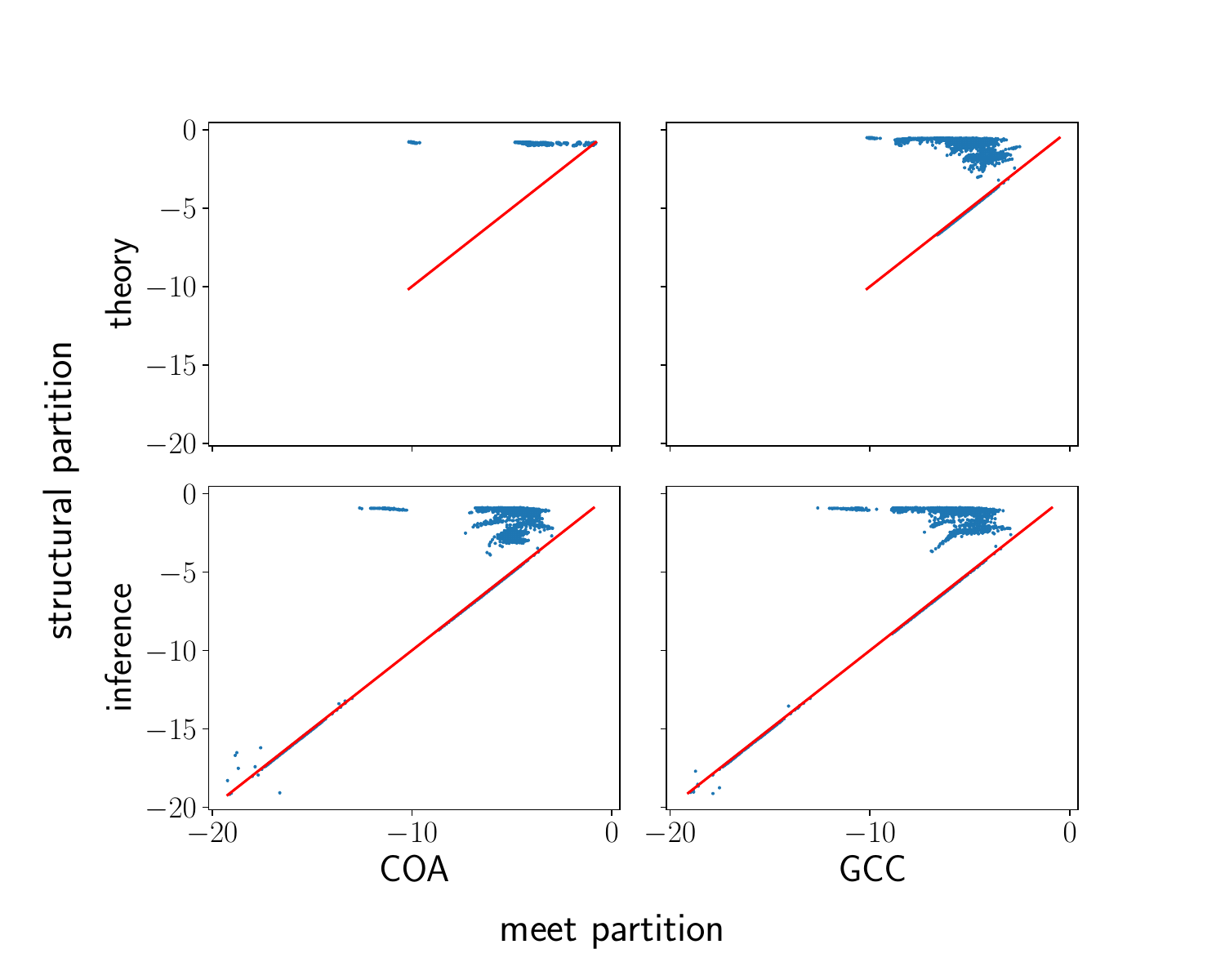} \vspace{-5ex}
    \caption{As in Fig. \ref{fig:log-rel-loss-theory_v_inference}, organized to show the effect of coarse-graining according to the meet partition vs. structural partition. Note that the meet partition is never worse than the structural one, and they agree whenever they are equal. This suggests that the added condition of phase-cohesiveness within modules (Condition \ref{ass:phase-cohesive-clusters}) does make a significant difference in one's ability to find an accurate coarse-grained model.}
    \label{fig:log-rel-loss-meet_v_structural}
\end{figure}

\begin{figure}
    \centering
    \includegraphics[width=\columnwidth]{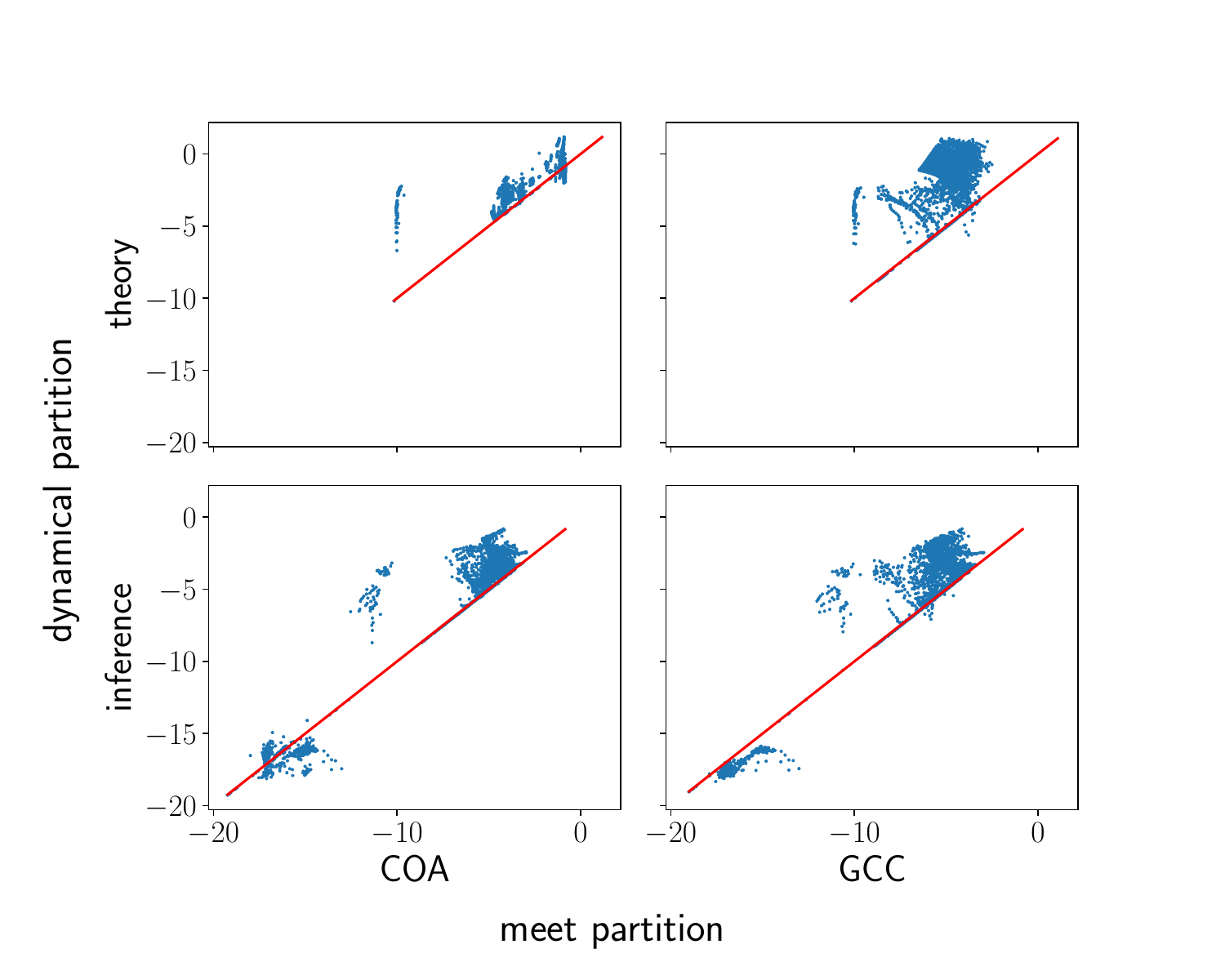} \vspace{-5ex}
    \caption{As in Fig. \ref{fig:log-rel-loss-theory_v_inference}, organized to show the effect of coarse-graining according to the meet partition vs. structural partition. Note that the meet partition only rarely outperforms the dynamical one, and they agree whenever they are equal. This suggests that the added condition of modularity of coupling strengths (Condition \ref{ass:modularity-of-coupling-matrix}) does make a significant difference in one's ability to find an accurate coarse-grained model.}
    \label{fig:log-loss-meet_v_dynamical}
    \vspace{5ex}
\end{figure}

\textbf{Quality of inferred models:}
In Figs. \ref{fig:log-rel-loss-theory_v_inference}--\ref{fig:log-loss-meet_v_dynamical} we take a more quantitative view on the results just described by forming scatter plots that make pairwise comparisons between coarse-grained models that differ in one design characteristic: partition, model class, and inference method. In each figure, every dot represents one instance of the coarse-graining problem that has been solved in each of twelve ways.

In Fig. \ref{fig:log-rel-loss-theory_v_inference} we examine the impact of inferring coefficients based on data vs. using their theoretical values. We see clearly that in all cases, inference affords better performance, but that in some cases theoretical coefficients perform nearly as well. This is the case for all situations except the COA anstaz using the structural partition, where the model with theoretical coefficients always performs poorly.

In Fig. \ref{fig:log-rel-loss-COA_v_GCC} we examine the impact of using the COA vs. the GCC ansatz. As noted qualitatively above, when allowing for inferred coefficient values, the two ans\"{a}tze attain comparable performance, while GCC is generally better when using theoretical values. This reflects the fact that the two function classes are equivalently expressive for the dynamics at hand, while the assumptions underlying the derivation of GCC are more closely reflective of the system under study than those leading to COA.

In Figure \ref{fig:log-rel-loss-meet_v_structural} we investigate the difference in coarse-grained model quality when coarse-graining according to the meet partition vs. the structural partition. This amounts to imposing Conditions \ref{ass:modularity-of-coupling-matrix} and \ref{ass:phase-cohesive-clusters} vs. only imposing Condition \ref{ass:modularity-of-coupling-matrix}. Unsurprisingly, the meet partition never gives poorer fit quality than the structural one, but there are cases along the diagonal when the two partitions coincide.

Finally, Fig. \ref{fig:log-loss-meet_v_dynamical} compares the meet partition with the dynamical partition. The meet partition generally outperforms the dynamical partition, indicating that Condition \ref{ass:modularity-of-coupling-matrix} is indeed necessary above and beyond Condition \ref{ass:phase-cohesive-clusters} to obtain a satisfactory coarse-grained model, even if one is allowed to optimize coefficients to fit the data. This suggests that mere phase-cohesiveness is not enough to justify a collective coordinate ansatz such as Eq. (\ref{eq:collective_coord_ansatz}). Rather, the cluster members should be influenced by other oscillators in the same manner -- have the same in-neighborhoods -- rendering their natural frequencies the only remaining differences between them to explain their different phases.

\subsection{Discussion and Path Forward} \label{sec:discussion}

In this study,
we introduced and numerically validated a data-driven procedure for discovering the reduced-order equations for the Kuramoto system in the presence of modular synchronization. In particular, we have demonstrated that in a modularly synchronized Kuramoto oscillator system, groups of mutually synchronized oscillators may be treated each as meta-oscillators, thereby yielding dynamics that belong to a family that strictly generalizes the Kuramoto model.

A central challenge of the problem considered here is to remove explicit time-dependence from the trivially-true coarse-grained model \eqref{eq:reduced_eqn_terms_gathered}. Here, we accomplished this task by imposing a set of assumptions on the coarse-graining map that justify a collective coordinate ansatz. An alternative approach requiring less advance knowledge of the underlying system would be to search over the space of partitions, either by brute-force or greedily, seeking those for which the coefficients \eqref{eq:A}--\eqref{eq:G} are a single-valued function of the coarse-grained state $\{z_\sigma\}$. If such a function exists, its substitution into Eq. \eqref{eq:reduced_eqn_terms_gathered} yields an autonomous ODE for the coarse variables.

Stripping away yet another layer of theory, we can remain largely agnostic about the functional form of the equations that should govern $\dot{z}$ and obtain the set basis functions using a procedure such as SINDy \cite{Brunton2015}, or use black-box representations such as neural-networks to discover a parsimonious basis. Still, given the enormous complexity of the space of functions $f\colon \mathbb{C}^C \to \mathbb{C}^C$, it is prudent to impose certain domain-informed restrictions on the function dictionary we choose.

First, the OA ansatz and its associated dimension reduction shows that the reduced model exhibits on-site and pairwise terms, and no higher-order (i.e. three-or-more-particle) terms. Moreover, each coupling term is of the same form. Therefore it is reasonable to suppose that our function basis should contain a set of onsite terms and a set of coupling terms, and that these terms should be replicated for each node and edge, respectively.

Next, it is known that coarse-graining of dynamics can induce memory effects \cite{Zwanzig1961}, so it is reasonable to expect that a coarse-grained model including memory or nonlocality (in the form of polyadic coupling terms) would be effective. Data-driven discovery of coarse-grained dynamical models including memory is discussed in \cite{Pan2018}. Polyadic coupling terms make no conceptual difference, but combinatorial explosion may make their inclusion computationally costly. A key potential advantage to including memory is the possibility of obtaining a good coarse-grained model with fewer state variables (i.e. a coarser partition).

Finally we remark on some physical principles that inform our methods for inferring coarse-grained dynamical models. Notice that in Eq. \eqref{eq:COA}, the coefficient of the linear term has negative real part (because $\delta_\sigma \ge 0$ is the width of a Cauchy distribution, and so cannot be negative), and the coupling matrix $K_{\sigma \sigma'}$ is real. These conditions together imply that the $C$-fold product of the unit disk is invariant; in other words, if all complex phases $z_\sigma$ initially have magnitude not greater than 1, then they will continue to have magnitude not greater than 1 for all time. This condition breaks down if we allow the coupling matrix to have an imaginary part or the coefficient of the linear term to have positive real part, and so {we impose those constraints during our inference procedure}.

Constraints of this type can be thought of in a more general context. We know that the coarse-grained variables $z_\sigma$ should always remain bounded in the unit disk, because they are averages of quantities on the unit circle. Knowing this, we perform \emph{constrained} inference of the right-hand side of an ODE governing those variables, with the constraint being that a function is feasible only if the corresponding ODE leaves the appropriate set in state space invariant. Fortunately, this constraint is convex, and is therefore straightforward to impose on the inference procedure.

\section{Conclusions} \label{sec:conclusion}

In this study we present a data-driven and theoretically-grounded approach to constructing coarse-grained models for finite-dimensional Kuramoto oscillator systems. To do this, we build on prior work that leverages the infinite-$N$ limit, as well as collective coordinate methods that offer the possibility of closing the explicit time-dependence induced by coarse-graining. We complement these theoretical approaches with a data-driven inference procedure that selects both the appropriate coarse-graining partition and optimal coefficient values for a given finite Kuramoto system. Our aim is to demonstrate the utility of combining theoretical insight with data-driven inference for finding accurate coarse-grained models of finite heterogeneous systems.

Our results reveal that partitioning nodes in a way that respects the underlying coupling network is vital to the validity of a collective coordinate approach, above and beyond phase cohesiveness. In addition, we find that given the freedom to optimize coefficient values based on data, the functional form predicted by Ott and Antonsen based on the assumption of Cauchy-distributed natural frequencies performs very comparably with the functional form predicted by the more realistic (for the case considered here) assumptions of Gaussian-distributed natural frequencies and a linear collective coordinate ansatz. This shows that there is considerable freedom in learning the right-hand side of a coarse-grained ODE, and that other criteria could be brought into play to disambiguate between comparable models. In the present work, the derivations given in Sec. \ref{sec:methods} perform this function in their role as inductive biases; other criteria such as parsimony \cite{Brunton2015} or interpretability might also be applied.

{
Notably, our method achieves accurate coarse-grained model performance across the whole parameter space we considered, which includes behaviors ranging from complete incoherence to global synchronization. The number of samples required to learn the coarse-grained model can vary according to the state of the system, and may drastically change when the system undergoes an abrupt phase transition \cite{Giannuzzi2007,gomez2011explosive}. In future work, it would be interesting to investigate the change of model complexity under phase transitions in original and coarse-grained systems.

Finally we discuss possible extensions to the work presented here. The derivations presented in Sec. \ref{subsec:assumptions} rely on partitioning nodes according to mathematical regularities in the underlying equations of motion, in particular regarding the coupling matrix. A similar approach could be applied in the case that the underlying system exhibits regularities in other properties, such as phase lags or higher harmonics in the coupling function, or in the on-site dynamics. We anticipate that the approach introduced in this work will find applications in other types of complex dynamical systems.
}

\section{Acknowledgements} \label{sec:ack}

This study was carried out at Los Alamos National Laboratory under the auspices of the National Nuclear Security Administration of the U.S. Department of Energy under Contract No. 89233218CNA000001. J.S. acknowledges support from Air Force Office of Scientific Research MURI FA9550-19-1-0386, US Army Research Office MURI Award No. W911NF-13-1-0340 and Cooperative Agreement No. W911NF-09-2-0053, and the U.S. Department of Defense, Minerva Grant No. W911NF-15-1-00502. A.Z. acknowledges support from the US Department of Energy National Nuclear Security Administration’s Office of Defense Nuclear Nonproliferation Research \& Development. A.Y.L. acknowledges support from the Laboratory Directed Research and Development program of Los Alamos National Laboratory under Projects No. 20190059DR, No. 20200121ER, and No. 20210078DR. This manuscript has been approved for unlimited release with control number LA-UR-20-27371.


\begin{thebibliography}{10}

\bibitem{Reynolds1895}
Osborne Reynolds.
\newblock {IV. On the dynamical theory of incompressible viscous fluids and the
  determination of the criterion}.
\newblock {\em Philosophical Transactions of the Royal Society of London.
  (A.)}, 186:123--164, 12 1895.

\bibitem{Chou1945}
P.~Y. Chou.
\newblock {On velocity correlations and the solutions of the equations of
  turbulent fluctuation}.
\newblock {\em Quarterly of Applied Mathematics}, 3(1):38--54, 4 1945.

\bibitem{Pope1975}
S.~B. Pope.
\newblock {A more general effective-viscosity hypothesis}.
\newblock {\em Journal of Fluid Mechanics}, 72(02):331, 11 1975.

\bibitem{Duraisamy2019}
Karthik Duraisamy, Gianluca Iaccarino, and Heng Xiao.
\newblock {Turbulence modeling in the age of data}.
\newblock {\em Annual Review of Fluid Mechanics}, 51:357--377, 2019.

\bibitem{Foley2015}
Thomas~T. Foley, M.~Scott Shell, and W.~G. Noid.
\newblock {The impact of resolution upon entropy and information in
  coarse-grained models}.
\newblock {\em Journal of Chemical Physics}, 143(24), 2015.

\bibitem{Boninsegna2018}
Lorenzo Boninsegna, Ralf Banisch, and Cecilia Clementi.
\newblock {A Data-Driven Perspective on the Hierarchical Assembly of Molecular
  Structures}.
\newblock {\em Journal of Chemical Theory and Computation}, 14(1):453--460,
  2018.

\bibitem{Wang2019}
Jiang Wang, Simon Olsson, Christoph Wehmeyer, Adrià P{\'{e}}rez, Nicholas~E.
  Charron, Gianni De~Fabritiis, Frank No{\'{e}}, and Cecilia Clementi.
\newblock {Machine Learning of Coarse-Grained Molecular Dynamics Force Fields}.
\newblock {\em ACS Central Science}, 2019.

\bibitem{Gleeson2012}
James~P. Gleeson, Sergey Melnik, Jonathan~A. Ward, Mason~A. Porter, and
  Peter~J. Mucha.
\newblock {Accuracy of mean-field theory for dynamics on real-world networks}.
\newblock {\em Physical Review E - Statistical, Nonlinear, and Soft Matter
  Physics}, 85(2):1--10, 2012.

\bibitem{Porter2016}
Mason~A. Porter and James Gleeson.
\newblock {\em {Dynamical Systems on Networks}}, volume~4 of {\em Frontiers in
  Applied Dynamical Systems: Reviews and Tutorials}.
\newblock Springer International Publishing, Cham, 2016.

\bibitem{kuramoto1975self}
Yoshiki Kuramoto.
\newblock {Self-entrainment of a population of coupled non-linear oscillators}.
\newblock In {\em International symposium on mathematical problems in
  theoretical physics}, volume~39, pages 420--422, Berlin/Heidelberg, 1975.
  Springer, Springer-Verlag.

\bibitem{aschoff1981circadian}
Jürgen Aschoff and Rütger Wever.
\newblock {The circadian system of man}.
\newblock In {\em Biological rhythms}, pages 311--331. Springer, 1981.

\bibitem{Strogatz1989}
Steven~H. Strogatz, Charles~M. Marcus, Robert~M. Westervelt, and Renato~E.
  Mirollo.
\newblock {Collective dynamics of coupled oscillators with random pinning}.
\newblock {\em Physica D: Nonlinear Phenomena}, 36(1-2):23--50, 1989.

\bibitem{Frank2000}
T.D. Frank, A.~Daffertshofer, C.E. Peper, P.J. Beek, and H.~Haken.
\newblock {Towards a comprehensive theory of brain activity: coupled oscillator
  systems under external forces}.
\newblock {\em Physica D: Nonlinear Phenomena}, 144(1-2):62--86, 9 2000.

\bibitem{Arenas2006}
Alexandre Arenas, Albert D{\'{i}}az-Guilera, and Conrad~J. P{\'{e}}rez-Vicente.
\newblock {Synchronization reveals topological scales in complex networks}.
\newblock {\em Physical Review Letters}, 96(11):1--4, 2006.

\bibitem{Li2008}
D.~Li, I.~Leyva, J.~A. Almendral, I.~Sendi{\~{n}}a-Nadal, J.~M. Buld{\'{u}},
  S.~Havlin, and S.~Boccaletti.
\newblock {Synchronization interfaces and overlapping communities in complex
  networks}.
\newblock {\em Physical Review Letters}, 101(16):2--5, 2008.

\bibitem{strogatz2000kuramoto}
Steven~H. Strogatz.
\newblock {From Kuramoto to Crawford: exploring the onset of synchronization in
  populations of coupled oscillators}.
\newblock {\em Physica D: Nonlinear Phenomena}, 143(1):1--20, 9 2000.

\bibitem{ott2008low}
Edward Ott and Thomas~M. Antonsen.
\newblock {Low dimensional behavior of large systems of globally coupled
  oscillators}.
\newblock {\em Chaos: An Interdisciplinary Journal of Nonlinear Science},
  18(3):037113, 9 2008.

\bibitem{Panaggio2015}
Mark~J Panaggio and Daniel~M. Abrams.
\newblock {Chimera states: coexistence of coherence and incoherence in networks
  of coupled oscillators}.
\newblock {\em Nonlinearity}, 28(3):R67--R87, 2015.

\bibitem{Bick2018}
Christian Bick, Mark~J Panaggio, and Erik~A. Martens.
\newblock {Chaos in Kuramoto oscillator networks}.
\newblock {\em Chaos: An Interdisciplinary Journal of Nonlinear Science},
  28(7):071102, 7 2018.

\bibitem{Zhang2019b}
Yuanzhao Zhang, Zachary~G. Nicolaou, Joseph~D. Hart, Rajarshi Roy, and
  Adilson~E. Motter.
\newblock {Critical Switching in Globally Attractive Chimeras}.
\newblock {\em Physical Review X}, 10(1):11044, 2019.

\bibitem{wiesenfeld1998frequency}
Kurt Wiesenfeld, Pere Colet, and Steven~H Strogatz.
\newblock Frequency locking in josephson arrays: Connection with the kuramoto
  model.
\newblock {\em Physical Review E}, 57(2):1563, 1998.

\bibitem{varela2001brainweb}
Francisco Varela, Jean-Philippe Lachaux, Eugenio Rodriguez, and Jacques
  Martinerie.
\newblock The brainweb: phase synchronization and large-scale integration.
\newblock {\em Nature reviews neuroscience}, 2(4):229--239, 2001.

\bibitem{kiss2002emerging}
Istv{\'a}n~Z Kiss, Yumei Zhai, and John~L Hudson.
\newblock Emerging coherence in a population of chemical oscillators.
\newblock {\em Science}, 296(5573):1676--1678, 2002.

\bibitem{zlotnik2016phase}
Anatoly Zlotnik, Raphael Nagao, István~Z. Kiss, and Jr-Shin Li.
\newblock {Phase-selective entrainment of nonlinear oscillator ensembles}.
\newblock {\em Nature communications}, 7(1):1--7, 2016.

\bibitem{dorfler2013synchronization}
Florian D{\"o}rfler, Michael Chertkov, and Francesco Bullo.
\newblock Synchronization in complex oscillator networks and smart grids.
\newblock {\em Proceedings of the National Academy of Sciences},
  110(6):2005--2010, 2013.

\bibitem{Gfeller2008}
David Gfeller and Paolo De~Los~Rios.
\newblock {Spectral coarse graining and synchronization in oscillator
  networks}.
\newblock {\em Physical Review Letters}, 100(17):1--4, 2008.

\bibitem{Izumida2013}
Yuki Izumida and Hiroshi Kori.
\newblock {Coarse-grained description of general oscillator networks}.
\newblock {\em arXiv}, pages 1--8, 2013.

\bibitem{Moon2006}
Sung~Joon Moon, R.~Ghanem, and Ioannis~G. Kevrekidis.
\newblock {Coarse graining the dynamics of coupled oscillators}.
\newblock {\em Physical Review Letters}, 96(14):1--4, 2006.

\bibitem{Rajendran2011}
Karthikeyan Rajendran and Ioannis~G. Kevrekidis.
\newblock {Coarse graining the dynamics of heterogeneous oscillators in
  networks with spectral gaps}.
\newblock {\em Physical Review E - Statistical, Nonlinear, and Soft Matter
  Physics}, 84(3):1--11, 2011.

\bibitem{Thiem2020}
Thomas~N. Thiem, Mahdi Kooshkbaghi, Tom Bertalan, Carlo~R. Laing, and
  Ioannis~G. Kevrekidis.
\newblock {Emergent spaces for coupled oscillators}.
\newblock {\em arXiv}, pages 1--31, 2020.

\bibitem{Gottwald2015}
Georg~A. Gottwald.
\newblock {Model reduction for networks of coupled oscillators}.
\newblock {\em Chaos}, 25(5), 2015.

\bibitem{Hancock2018}
Edward~J. Hancock and Georg~A. Gottwald.
\newblock {Model reduction for Kuramoto models with complex topologies}.
\newblock {\em Physical Review E}, 98(1):012307, 7 2018.

\bibitem{Smith2019}
Lachlan~D. Smith and Georg~A. Gottwald.
\newblock {Chaos in networks of coupled oscillators with multimodal natural
  frequency distributions}.
\newblock {\em Chaos: An Interdisciplinary Journal of Nonlinear Science},
  29(9):093127, 2019.

\bibitem{Smith2019a}
Jan~R. Engelbrecht and Renato Mirollo.
\newblock {Is the Ott-Antonsen manifold attracting?}
\newblock {\em Physical Review Research}, 2(2):023057, 4 2020.

\bibitem{Yue2020}
Wenqi Yue, Lachlan~D. Smith, and Georg~A. Gottwald.
\newblock {Model Reduction for the Kuramoto-Sakaguchi Model: The Importance of
  Non-entrained Rogue Oscillators}.
\newblock {\em arXiv}, pages 1--14, 2020.

\bibitem{Kuramoto1984}
Yoshiki Kuramoto.
\newblock {Cooperative Dynamics of Oscillator Community}.
\newblock {\em Progress of Theoretical Physics Supplement}, 79(0):223--240,
  1984.

\bibitem{Skardal2012}
Per~Sebastian Skardal and Juan~G. Restrepo.
\newblock {Hierarchical synchrony of phase oscillators in modular networks}.
\newblock {\em Physical Review E - Statistical, Nonlinear, and Soft Matter
  Physics}, 85(1):1--8, 2012.

\bibitem{Schaub2016}
Michael~T. Schaub, Neave O'Clery, Yazan~N. Billeh, Jean-Charles~Charles
  Delvenne, Renaud Lambiotte, and Mauricio Barahona.
\newblock {Graph partitions and cluster synchronization in networks of
  oscillators}.
\newblock {\em Chaos}, 26(9):1--38, 2016.

\bibitem{Pecora2014}
Louis~M. Pecora, Francesco Sorrentino, Aaron~M. Hagerstrom, Thomas~E. Murphy,
  and Rajarshi Roy.
\newblock {Cluster synchronization and isolated desynchronization in complex
  networks with symmetries.}
\newblock {\em Nature communications}, 5(May):4079, 2014.

\bibitem{Sorrentino2015}
Francesco Sorrentino, Louis~M. Pecora, Aaron~M. Hagerstrom, Thomas~E. Murphy,
  and Rajarshi Roy.
\newblock {Complete characterization of the stability of cluster
  synchronization in complex dynamical networks}.
\newblock {\em Science Advances}, 2(4):e1501737, 4 2016.

\bibitem{Brunton2015}
Steven~L. Brunton, Joshua~L. Proctor, and J.~Nathan Kutz.
\newblock {Discovering governing equations from data by sparse identification
  of nonlinear dynamical systems}.
\newblock {\em Proceedings of the National Academy of Sciences},
  113(15):3932--3937, 4 2016.

\bibitem{lokhov2018online}
Andrey~Y Lokhov, Marc Vuffray, Dmitry Shemetov, Deepjyoti Deka, and Michael
  Chertkov.
\newblock Online learning of power transmission dynamics.
\newblock In {\em 2018 Power Systems Computation Conference (PSCC)}, pages
  1--7. IEEE, 2018.

\bibitem{Bialonski2006}
Stephan Bialonski and Klaus Lehnertz.
\newblock {Identifying phase synchronization clusters in spatially extended
  dynamical systems}.
\newblock {\em Physical Review E - Statistical, Nonlinear, and Soft Matter
  Physics}, 74(5):1--11, 2006.

\bibitem{Groth2011}
Andreas Groth and Michael Ghil.
\newblock {Multivariate singular spectrum analysis and the road to phase
  synchronization}.
\newblock {\em Physical Review E - Statistical, Nonlinear, and Soft Matter
  Physics}, 84(3):1--10, 2011.

\bibitem{Allefeld2007}
Carsten Allefeld, Markus M{\"{u}}ller, and Jürgen Kurths.
\newblock {Eigenvalue decomposition as a generalized synchronization cluster
  analysis}.
\newblock {\em International Journal of Bifurcation and Chaos},
  17(10):3493--3497, 2007.

\bibitem{Virtanen2020}
Pauli Virtanen, Ralf Gommers, Travis~E. Oliphant, Matt Haberland, Tyler Reddy,
  David Cournapeau, Evgeni Burovski, Pearu Peterson, Warren Weckesser, Jonathan
  Bright, Stéfan~J. van~der Walt, Matthew Brett, Joshua Wilson, K.~Jarrod
  Millman, Nikolay Mayorov, Andrew R.~J. Nelson, Eric Jones, Robert Kern, Eric
  Larson, C~J Carey, İlhan Polat, Yu~Feng, Eric~W. Moore, Jake VanderPlas,
  Denis Laxalde, Josef Perktold, Robert Cimrman, Ian Henriksen, E.~A. Quintero,
  Charles~R. Harris, Anne~M. Archibald, Antônio~H. Ribeiro, Fabian Pedregosa,
  and Paul van Mulbregt.
\newblock {SciPy 1.0: fundamental algorithms for scientific computing in
  Python}.
\newblock {\em Nature Methods}, 17(3):261--272, 3 2020.

\bibitem{Zwanzig1961}
Robert Zwanzig.
\newblock {Memory effects in irreversible thermodynamics}.
\newblock {\em Physical Review}, 124(4):983--992, 1961.

\bibitem{Pan2018}
Shaowu Pan and Karthik Duraisamy.
\newblock {Data-Driven Discovery of Closure Models}.
\newblock {\em SIAM Journal on Applied Dynamical Systems}, 17(4):2381--2413, 1
  2018.

\bibitem{Giannuzzi2007}
F.~Giannuzzi, D.~Marinazzo, G.~Nardulli, M.~Pellicoro, and S.~Stramaglia.
\newblock {Phase diagram of a generalized Winfree model}.
\newblock {\em Physical Review E - Statistical, Nonlinear, and Soft Matter
  Physics}, 75(5):1--6, 2007.

\bibitem{gomez2011explosive}
Jesús G{\'{o}}mez-Garde{\~{n}}es, Sergio G{\'{o}}mez, Alexandre Arenas, and
  Yamir Moreno.
\newblock {Explosive synchronization transitions in scale-free networks}.
\newblock {\em Physical review letters}, 106(12):128701, 2011.

\end{thebibliography}
\end{document}